\documentclass[reqno,11pt]{amsart}
\usepackage{amsfonts}
\usepackage{graphicx}
\usepackage[mathscr]{euscript}
\numberwithin{equation}{section}
\allowdisplaybreaks[1]

\title[Causal Variational Principles on Measure Spaces]{Causal Variational Principles on Measure Spaces}

\author[F.\ Finster]{Felix Finster \\ \\ November 2008}
\thanks{Supported in part by the Deutsche Forschungsgemeinschaft.}
\address{Fakult\"at f\"ur Mathematik \\ Universit\"at Regensburg \\ D-93040 Regensburg \\ Germany}
\email{Felix.Finster@mathematik.uni-regensburg.de}

\newtheorem{Def}{Definition}[section]
\newtheorem{Thm}[Def]{Theorem}

\newtheorem{Lemma}[Def]{Lemma}

\newtheorem{Corollary}[Def]{Corollary}
\newtheorem{Example}[Def]{Example}

\newcommand{\Thanks}{\vspace*{.5em} \noindent \thanks}
\newcommand{\beq}{\begin{equation}}
\newcommand{\eeq}{\end{equation}}
\newcommand{\Proof}{\begin{proof}}
\newcommand{\QED}{\end{proof} \noindent}
\newcommand{\QEDrem}{\ \hfill $\Diamond$}

\DeclareMathOperator{\lbra}{\langle}
\DeclareMathOperator{\lket}{\rangle}
\DeclareMathOperator{\Sl}{\prec\!}
\DeclareMathOperator{\Sr}{\!\succ}
\DeclareMathOperator{\bra}{<\!}
\DeclareMathOperator{\ket}{\!>}
\newcommand{\C}{\mathbb{C}}
\newcommand{\R}{\mathbb{R}}
\newcommand{\1}{\mbox{\rm 1 \hspace{-1.05 em} 1}}

\newcommand{\N}{\mathbb{N}}

\DeclareMathOperator{\Tr}{Tr}
\DeclareMathOperator{\tr}{tr}
\DeclareMathOperator{\Mat}{Mat}

\DeclareMathOperator{\diam}{diam}
\renewcommand{\O}{{\mathscr{O}}}
\renewcommand{\L}{{\mathcal{L}}}
\newcommand{\Lin}{\text{\rm{L}}}
\newcommand{\Sact}{{\mathcal{S}}}
\newcommand{\T}{{\mathcal{T}}}

\newcommand{\F}{{\mathscr{F}}}
\newcommand{\M}{{\mathbb{M}}}
\newcommand{\U}{\text{\rm{U}}}
\newcommand{\Ub}{{\mathbb{U}}}
\newcommand{\B}{{\mathcal{B}}}
\newcommand{\K}{{\mathcal{K}}}
\newcommand{\m}{{\mathfrak{m}}}
\newcommand{\n}{{\mathfrak{n}}}
\renewcommand{\H}{{\mathscr{H}}}

\newcommand{\e}{{\mathfrak{e}}}
\newcommand{\f}{{\mathfrak{f}}}

\begin{document}
\maketitle

\begin{abstract}
We introduce a class of variational principles on measure spaces
which are causal in the sense that they generate a
relation on pairs of points, giving rise to a
distinction between spacelike and timelike separation.
General existence results are proved. It is shown in examples that minimizers
need not be unique. Counter examples to compactness are discussed.
The existence results are applied to variational principles
formulated in indefinite inner product spaces.
\end{abstract}

\tableofcontents
Causal variational principles on measure spaces arise in the
context of relativistic quantum theory.
But they are also interesting from a purely mathematical perspective
as a class of nonlinear variational principles whose minimizers have a surprisingly rich
and so far largely unexplored structure.
The goal of the present article is to give a mathematical introduction to these variational principles
and to develop the existence theory (Chapters~\ref{sec1} and~\ref{sec2}).
The physical applications will be obtained by reformulating the variational principles
in indefinite inner product spaces (Chapters~\ref{sec3} and~\ref{sec4}).
Our results will be illustrated by a number of examples and counter examples, which also
show that the minimizers are in general non-trivial and not unique.

\section{The Causal Variational Principle with Two Prescribed Eigenvalues} \label{sec1}
\subsection{Introduction and Basic Definitions}
In order to give an easily accessible introduction to the basic ideas and methods,
we begin with the simplest interesting example, the so-called causal variational principle
with two prescribed eigenvalues (the general systems will be introduced in Chapter~\ref{sec2}).
Let~$(M, \mu)$ be a measure space of total volume~$\mu(M)=1$.
For a given parameter $\beta \in [0,1)$ and an integer~$f \geq 2$, we let $\F \subset
\Mat(\C^f)$ be the set of all Hermitian $f \times f$-matrices of rank at most two, whose
non-trivial eigenvalues are equal to~$1$ and~$-\beta$.
Consider the set of matrix-valued functions
\[ \M = \left\{ F : M \rightarrow \F \text{ measurable} \right\} . \]
For any~$F \in \M$ and~$x, y \in M$, the matrix product
\beq \label{Adef}
A_{xy} = F(x)\cdot F(y)
\eeq
is of rank at most two. Thus counting with algebraic multiplicities, its
eigenvalues are
\[ \lambda_+^{xy}, \lambda_-^{xy}, \underbrace{0,\ldots, 0}_{\text{$f-2$ times}}
\quad \text{with} \quad \lambda_\pm^{xy} \in \C \]
(note that the matrix~$A_{xy}$ is Hermitian only in the special case that~$F(x)$
and~$F(y)$ commute, and thus the~$\lambda_\pm^{xy}$ will in general be complex).
Clearly, the functions~$\lambda_\pm^{xy}$ are measurable in~$x$ and~$y$.
Thus introducing
\beq
\boxed{
\begin{split}
\text{the Lagrangian} \quad\quad \L[A_{xy}] &= \frac{1}{2} \left(|\lambda_+^{xy}|
- |\lambda_-^{xy}| \right)^2 \\
\text{the action} \qquad\quad\quad\! \Sact[F] &= \iint_{M \times M}  \L[A_{xy}] \, d\mu(x)\, d\mu(y)\:,
\end{split} } \label{LSdef}
\eeq
we obtain a non-negative functional~$\Sact$ on~$\M$. Our variational principle is to
\[ \text{minimize } \Sact \text{ on } \M\:. \]
We are interested in the following questions:
\begin{itemize}
\item What is the infimum of the action? Is the infimum attained?
\item Provided a minimizer exists, what is its regularity? Is the minimizer unique?
What is the structure of the minimizers?
\end{itemize}

Before addressing these questions, we explain the form of the Lagrangian and discuss a few
properties of our variational principle. We first point out that, since we prescribed its
eigenvalues, every matrix in~$\F$ has sup-norm one, and thus
\[ |\lambda^{xy}_\pm| \leq \|F(x) \,F(y)\| \leq \|F(x)\| \,\|F(y)\| = 1 \:. \]
Hence the Lagrangian is bounded, $\L \in L^\infty(M \times M, \R)$, and the
action is finite. Next, the transformations
\[ \lambda_+^{xy} + \lambda_-^{xy} = \Tr(F(x) F(y)) = \overline{ \Tr(F(y) F(x)) }
= \overline{\Tr(F(x) F(y))} = \overline{\lambda_+^{xy}} + \overline{\lambda_-^{xy} } \]
show that the~$\lambda_+^{xy}$ and~$\lambda_-^{xy}$ are either both real, or else they
form a complex conjugate pair. This distinction gives rise to a notion of causality.
\begin{Def} \label{def1} Two points~$x, y \in M$ are called {\bf{timelike}} and {\bf{spacelike}} separated if the roots~$\lambda_\pm^{xy}$ of the characteristic polynomial of~$A_{xy}$ are real
or non-real, respectively.
\end{Def} \noindent
Let us verify that this notion is symmetric in~$x$ and~$y$. The first method is to
note that the~$\lambda_\pm^{xy}$ are uniquely determined as the solutions of the two equations
\[ \lambda_+^{xy} + \lambda_-^{xy} = \Tr(A_{xy}) \qquad \text{and} \qquad
(\lambda_+^{xy})^2 + (\lambda_-^{xy})^2 = \Tr \left( A_{xy}^2 \right) . \]
Substituting~\eqref{Adef} and cyclically commuting the arguments of the traces, one sees that
the traces are invariant under exchanging~$x$ and~$y$. In other words, 
the matrices~$A_{xy}$ and~$A_{yx}$ have the same spectrum, showing that our notion of causality
is indeed symmetric in~$x$ and~$y$. Alternatively, this can be seen from the
matrix identity~$\det(BC-\lambda \1)=\det(CB-\lambda \1)$ (see for example~\cite[Section~3]{discrete}).

If~$x$ and~$y$ are spacelike separated, we just saw that the~$\lambda_\pm^{xy}$ form a
complex conjugate pair. Hence the absolute values of~$\lambda_+^{xy}$
and~$\lambda_-^{xy}$ coincide, and thus the Lagrangian~$\L$ in~\eqref{LSdef} vanishes.
In other words, pairs~$(x,y)$ with spacelike separation drop out of the Lagrangian and thus
do not enter the action.
We refer to this fact that our variational principle is {\em{causal}}.
This causality can be seen in analogy to relativity, where space-time points with spacelike
separation cannot influence each other via the physical equations. This analogy will become
clearer in Chapter~\ref{sec3}, when variational principles in indefinite inner product spaces are
considered. Until then, we shall focus on the mathematical properties of our variational principle.

Qualitatively, since the Lagrangian vanishes for spacelike separation, our variational principle
tries to achieve that as many pairs of points as possible have spacelike separation.
On the other hand, in the special case~$F(x)=F(y)$, the matrix~$A_{xy}$ has the non-zero
eigenvalues~$1$ and~$\beta^2$, showing that~$x$ and~$y$ will have timelike separation
if~$F(x)$ and~$F(y)$ are sufficiently close to each other. \label{strictlyp}
Thus there are competing mechanisms, and this will lead to mathematically interesting effects.

Another point of mathematical interest is that our variational principle
generates mathematical structures on~$M$. Note that~$(M, \mu)$ is merely a measure space, but we
do not assume a topology. But a given minimizer~$F$ induces a topology on~$M$
(namely~$F^{-1}(\O)$, where~$\O$ denotes the set of all open subsets of~$\F$), and furthermore~$F$ induces the causal structure of Definition~\ref{def1}. Thus a minimizer
{\em{generates on~$M$ a topological and causal structure}}. In order to better understand this
structure formation, one needs to clarify the freedom in choosing the
minimizers of the variational principle. In particular, if our variational principle allowed
to distinguish a specific minimizer~$F$ (determined modulo isomorphisms of the measure
space~$(M, \mu)$), this would give rise to a canonical topology and a canonical causal
structure on~$M$.

\subsection{Existence of Minimizers} \label{sec12}
This section is devoted to the proof of the following general existence theorem.
\begin{Thm} \label{thm1} There is a function~$F \in \M$ such that
\[ \Sact[F] = \inf_\M \Sact\:. \]
\end{Thm}
The most obvious idea for the proof is to try the direct method of the calculus of variations.
Thus let~$F_k \in \M$ be a minimizing sequence, i.e.
\[ \lim_{k \rightarrow \infty} \Sact[F_k] = \inf_\M \Sact\:. \]
The proof would be completed if we found a convergent subsequence~$F_{n_k}$
and could prove that~$\Sact$ was lower semi-continuous. The following consideration explains why
this method does not seem to work. If our subsequence converged in the
weak sense, $F_{n_k} \rightharpoondown F$, the spectral properties of the
matrices~$F_{n_k}(x)$ could not be controlled in the limit. Thus the matrix~$F(x)$ would in general
no longer have the eigenvalues~$1$ and~$-\beta$, and thus~$F$ would not be the desired minimizer.
This explains why for a useful notion of convergence in~$\F$ it seems necessary to consider
the topology induced by the $\sup$-distance function
\beq \label{dFG}
d(F,G) = \sup_{x \in M} \|F(x)-G(x)\|\:,
\eeq
where~$\|.\|$ is a matrix norm on~$\F$.
Now suppose that~$\phi$ is an isomorphism of the measure
space~$(M, \mu)$ (i.e.\ a measure preserving bijection of~$M$). Then for any given~$F_0 \in \M$, the
function~$F_0 \circ \phi$ is again in~$\M$, and both functions have the same action. More generally,
$\Sact$ is constant on the orbit~$\Ub$ of~$F_0$ under the action of such isomorphisms,
\beq \label{Udef}
\Ub := \left\{ F_0 \circ \phi \:|\: \text{$\phi$ isomorphism of~$(M, \mu)$} \right\} \subset \M\:.
\eeq
The problem is that the orbits~(\ref{Udef}) are in general not compact in the topology~(\ref{dFG}).
To see this in a simple example, we take~$M=[0,1)$ with the Lebesgue measure
and choose~$F_0 \in \M$ as a function which takes two different values~$p,q \in \F$, being constant on the
intervals~$[0,\frac{1}{2})$ and~$[\frac{1}{2}, 1)$. We
consider the one-parameter family of isomorphisms~$\phi_\lambda(x)=(x+\lambda) \bmod 1$
with~$\lambda \in [0,1)$.
Then the functions~$F_\lambda(x) := F_0 \circ \phi_\lambda$ are all in~$\Ub$, but
for any~$\lambda \neq \mu$ their distance is a non-zero constant, $d(F_\lambda, F_\mu)=d(p,q)>0$.
Hence there is even an uncountable family of functions in~$\Ub$ which has no
convergent subsequence.

Our method to avoid the above problem is to translate the functions~$F_k$ into
measures~$\rho_k$ on~$\F$, as we now explain. We first note that in the case~$0<\beta<1$,
every point $p \in \F$ is a $f \times f$-matrix, which is characterized by the two orthogonal
eigenspaces corresponding to the eigenvalues~$1$ and~$-\beta$.
Characterizing~$p$ equivalently by the first eigenspace and the linear span of both eigenspaces,
we can identify~$p$ with a point of the {\em{flag manifold}}~$\F^{1,2}(\C^f)$
(for the detailed definition we refer to~\cite[Chapter~I, \S3.1]{helgason}).
Likewise, in the case~$\beta=0$, every point~$p \in \F$ is characterized by the eigenspace
corresponding to the eigenvalue one, and thus~$\F$ can be identified with the Grassmannian~$\F^1(\C^f)$.
In each case, this identification is useful because it makes~$\F$ into a smooth compact manifold.
Moreover, $\F$ is a homogeneous space, meaning that the mapping
\beq \label{transitive}
p \rightarrow UpU^{-1} \qquad \text{with} \qquad U \in \U(f)
\eeq
defines a transitive action of the group~$\U(f)$ on~$\F$.
We introduce on~$\F$ a Riemannian metric~$g$ which is invariant under this group action
and denote the corresponding invariant measure by~$\mu_\F$. For simplicity, we normalize~$g$
such that~$\mu_\F(\F)=1$. Taking the infimum of the lengths of curves gives on~$\F$ a distance function
\[ d : \F \times \F \rightarrow \R^+_0\:. \]
The topology of~$\F$ is generated by the open balls~$B_\varepsilon(y)$
of distance radius~$\varepsilon$ centered at~$y \in \F$.

Next to any~$F \in \M$ we introduce a measure~$\rho$ on~$\F$ by defining that~$\Omega \subset \F$
is measurable if and only if~$F^{-1}(\Omega) \subset M$ is measurable and by setting
\beq \label{rhodef}
\rho(\Omega) = \mu(F^{-1}(\Omega))\:.
\eeq
Clearly, $\rho(\F) = \mu(M)=1$. The advantage of working with~$\rho$ is that it does
not depend on isomorphisms of~$(M, \mu)$, as the simple calculation
\[ \mu((F \circ \phi)^{-1}(\Omega)) = \mu(\phi^{-1}(F^{-1}(\Omega))) = 
\mu(F^{-1}(\Omega)) \]
shows. Furthermore, our action can be expressed in terms of the measure~$\rho$ by
\beq \label{Srho}
\Sact(\rho) = \iint_{\F \times \F} \L[p \cdot q]\: d\rho(p) \,d\rho(q) \:,
\eeq
where~$p \cdot q$ is the matrix multiplication of elements of~$\F$.
Our strategy is to first construct a minimizer of~\eqref{Srho} and then to
construct the corresponding minimizer~$F$ of the original variational problem.

Let~$C^0(\F)$ be the Banach space of continuous functions on~$\F$, equipped with the $\sup$-norm.
We consider every measure~$\rho_k$ according to~$\rho_k(f) = \int_\F f\, d\rho_k$ (with~$f \in C^0(\F)$)
as a linear functional on~$C^0(\F)$. The relations
\beq \label{rel1}
|\rho_k(f)| \leq \|f\| \, \rho_k(\F) = \|f\| \:,\qquad \rho_k(1_\F) = \rho_k(\F) = 1
\eeq
(where~$1_\F : \F \rightarrow \R$ is the constant function one)
yield that the~$\rho_k$ are continuous functionals and $\|\rho_k\|_{C^0(\F)^*} = 1$.
The positivity of the measures~$\rho_k$ is expressed by
\beq \label{rel2}
\rho_k(f) \geq 0 \quad \text{for all $f \in C^0(\F)$ with~$f \geq 0$}\:.
\eeq
The Banach-Alaoglu theorem~\cite{reed+simon}
yields a subsequence, for simplicity again denoted by~$\rho_k$,
which converges in the weak-*-topology; that is, for every~$f \in C^0(\F)$
the series $\rho_k(f)$ converges. Thus by~$\rho(f) = \lim_k \rho_k(f)$
we can define a functional on~$C^0(\F)$.
By the Riesz representation theorem~\cite{rudinFA},
there is a regular Borel measure~$\rho$ on~$\F$ such that
\[ \rho(f) = \int_{\F} f \,d\rho \qquad \text{for all~$f \in C^0(\F)$}\:. \]
From~(\ref{rel1}) one sees that~$\rho$ is normalized to~$\rho(\F)=1$.
Furthermore,  taking the limit~$k \rightarrow \infty$ in~(\ref{rel2}) one sees that~$\rho$
is a positive measure.
We conclude that there is a subsequence~$\rho_k$ and a positive normalized regular Borel measure such that
\[ \int_{\F} f \, d\rho_k \rightarrow \int_{\F} f \, d\rho \qquad \text{for all~$f \in C^0(\F)$}\:. \]
Since the function~$\L[p \cdot q]$ with~$\L$ according to~(\ref{LSdef}) is
continuous in both arguments~$p, q \in \F$, we conclude that~$\Sact(\rho_k) \rightarrow \Sact(\rho)$.
We have thus proved the following result.
\begin{Lemma} \label{lemmameasure} There is a positive normalized regular
Borel measure~$\rho$ on~$\F$ such that
\[ \iint_{\F \times \F} \L[p \cdot q]\: d\rho(p) \,d\rho(q) = \inf_{\F} S[F]\:. \]
\end{Lemma}

The remaining step is to ``realize'' the measure~$\rho$ by a function~$F \in \M$ as follows.
\begin{Lemma} \label{lemma14}
There is a measurable function~$F \in \M$ such that the measure~$\rho$
of Lemma~\ref{lemmameasure} has the representation
\[ \rho(\Omega) = \mu(F^{-1}(\Omega)) \quad \text{for every Borel set~$\Omega \subset \F$}\:. \]
\end{Lemma}
\Proof We recall that a measurable set~$\Omega \subset \M$ is called an {\em{atom}} if
$\mu(\Omega)>0$ and if every subset~$K \subset \Omega$ with~$\mu(K) < \mu(\Omega)$
has measure zero (cf.~\cite[Section~40]{halmosmt}). A measure is said to be
{\em{atomic}} if every set of non-zero measure contains an atom. Conversely, a measure is
{\em{non-atomic}} if it contains no atoms. The measure~$\mu$ can be decomposed into the sum of
an atomic measure~$(\mu^\text{d}, M^\text{d})$ and a non-atomic measure~$(\mu^\text{c}, M^\text{c})$ in the sense that
\beq \label{Mdec}
M = M^\text{d} \,\dot{\cup}\, M^\text{c} \qquad \text{and} \qquad \mu = \mu^\text{d} + \mu^\text{c}
\eeq
(see~\cite{johnson} for a proof in general measure spaces).
Furthermore, an exhaustion argument (see~\cite[Section~42 (2)]{halmosmt}) shows that
\beq \label{decomp}
\text{For all } \alpha \in [0, \mu^\text{c}(M^\text{c})] \text{ there is a measurable set }
\Omega \subset M^\text{c} \text{ with } \mu^\text{c}(\Omega)=\alpha\:.
\eeq

In order to treat the atomic part, we denote the atoms of the measure~$(\mu^\text{d}, M^\text{d})$
by~$\mathfrak{A}$. Since every element~$A \in \mathfrak{A}$ has measure~$\mu^\text{d}(A)>0$
and
\[ \sum_{A \in \mathfrak{A}} \mu^\text{d}(A) = \mu(M^\text{d}) \leq 1\:, \]
the set~$\mathfrak{A}$ is at most countable. Furthermore, by modifying the functions~$F_k$ on
sets of measure zero we can arrange that for every~$A \in \mathfrak{A}$,
the set~$F_k(A)$ consist of just one point, for simplicity again denoted by~$F_k(A)$.
For any~$A \in \mathfrak{A}$ we consider the series~$F_k(A) \in \F$.
Since~$\F$ is compact, this series has an accumulation point.
Using that~$\mathfrak{A}$ is at most countable, a diagonal series argument
yields that, possibly after choosing
again a subsequence of~$(F_k)_{k \in \N}$, for every~$A \in \mathfrak{A}$
the series~$F_k(A)$ converges in~$\F$ as~$k \rightarrow \infty$.
Setting~$f(A) = \lim_{k \rightarrow \infty} F_k(A)$, it follows that for
every~$\varepsilon>0$ and~$y \in \F$,
\[ \rho(B_\varepsilon(y)) = 
\lim_{k \rightarrow \infty} \rho_k(B_\varepsilon(y)) \;\geq\; \sum_{A \in \mathfrak{A} \text{ with }
f(A)=y} \mu^\text{d}(A) \:. \]
Since~$\rho$ is a regular Borel measure, we can let~$\varepsilon \rightarrow 0$
to obtain the same bound for~$\rho(\{y\})$. Hence subtracting a sum of Dirac measures,
\beq \label{rhocdef}
\rho^\text{c} = \rho - \sum_{A \in \mathfrak{A}} \mu^\text{d}(A) \:\delta_{f(A)}\,,
\eeq
the resulting measure~$\rho^\text{c}$ on~$\F$ is again positive. Defining the function~$F$
on~$M^\text{d}$ by
\[ F|_{M^\text{d}} \::\: M^\text{d} \rightarrow \F \::\: A \mapsto f(A)\:, \]
it remains to construct a function~$g:=F|_{M^\text{c}} \,:\, M^\text{c} \rightarrow \F$ such that
\beq \label{gcond}
\rho^\text{c}(\Omega) = \mu^\text{c}(g^{-1}(\Omega))  \quad
\text{for every Borel set~$\Omega \subset \F$}\:.
\eeq

To handle the non-atomic measure~$\mu^\text{c}$, we proceed  by induction in~$k$.
In the first step~$k=1$, we decompose~$\F$ into a disjoint finite union of non-empty Borel sets of
bounded distance diameter,
\[ \F = \F^1_1 \dot{\cup} \cdots \dot{\cup} \F^1_{l_{\max}(1)} \qquad \text{with} \qquad
\diam(\F^1_l) < 1 . \]
Since~$\mu^\text{c}$ is non-atomic and
\[ \mu^\text{c}(M^\text{c}) = 1 - \sum_{A \in \mathfrak{A}} \mu^\text{d}(A)
\overset{\eqref{rhocdef}}{=} \rho^\text{c}(\F)\:, \]
we can apply~\eqref{decomp} iteratively to decompose~$M^\text{c}$ into a disjoint union of
measurable sets~$M^1_1, \ldots, M^1_{l_{\max}(1)}$
with~$\mu^\text{c}(M^1_l) = \rho^\text{c}(F^1_l)$.
Of every set~$\F^1_l$ we choose a point~$y^1_l$ and define a step function~$g_1 \,:\, M^\text{c} \rightarrow \F$ by~$g_1|_{M^1_l} \equiv y^1_l$, $l=1,\ldots, l_{\max}(1)$.

For the iteration step~$(k-1) \rightarrow k$ we decompose each of the sets~$\F^{k-1}_1, \ldots, \F^{k-1}_{l_{\max}(k-1)}$ into sets of smaller diameter to obtain a decomposition of~$\F$ of the form
\[ \F = \F^k_1 \dot{\cup} \cdots \dot{\cup} \F^k_{l_{\max}(k)} \qquad \text{with} \qquad
\diam(\F^k_l) < \frac{1}{k} . \]
Again using~\eqref{decomp}, we subdivide the sets~$M^{k-1}_l$ into smaller
sets~$M^k_1, \ldots, M^k_{l_{\max}(k)}$ with the
property~$\mu^\text{c}(M^k_l) = \rho^\text{c}(\F^k_l)$.
Of every set~$\F^k_l$ we again choose a point~$y^k_l$ and define a step function~$g_k \,:\, M^\text{d} \rightarrow \F$ by~$g_k|_{M^k_l} \equiv y^k_l$, $l=1,\ldots, l_{\max}(k)$.

This inductive procedure gives a series of step functions~$(g_k)_{k \in \N}$.
By construction, this series converges uniformly to a measurable
function~$g \,:\, M^\text{d} \rightarrow \F$. Furthermore, for any Borel set~$\Omega \subset \F$,
\[ \mu^\text{c}(g^{-1}(\Omega)) \xleftarrow{k \rightarrow \infty}
\sum_{y^k_l \in \Omega} \mu^\text{c}(M_l^k) = 
\sum_{y^k_l \in \Omega} \rho^\text{c}(\F_l^k)
\xrightarrow{k \rightarrow \infty} \rho^\text{c}(\Omega)\:. \]
Thus~$g$ satisfies~\eqref{gcond}, completing the proof.
\QED

This completes the proof of Theorem~\ref{thm1}.

\subsection{Non-Uniqueness in the Discrete Setting}
We now discuss the uniqueness problem in the case when the measure~$\mu$ is discrete.
We begin with the simple example where~$M=\{1\}$ consists of only one point.
In this case, every function~$F \in \M$ can be identified with a point~$F(1) \in \F$.
All these functions give the same value~$\Sact = \frac{1}{2}(1-\beta^2)^2$ of the action,
and thus the minimizer is clearly not unique. However, since according to~\eqref{transitive}
the $\U(f)$-symmetry group acts transitively on~$\F$, all the minimizers can be obtained from
each other by a suitable $\U(f)$-transformation. Thus the minimizer is unique up to
$\U(f)$-transformations on~$\F$.

The next example gives a connection to a problem already studied in the literature
and gives a good intuition for the mechanisms in our variational principle.
\begin{Example} {{(Maximizing distances of points on~$S^2$)}} \label{exPauli}
\em We choose~$f=2$, $\beta=0$ and let~$M=\{1, \ldots, m\}$ be a finite set with the
normalized counting measure $\mu(\{1\})=\cdots=\mu(\{m\})=\frac{1}{m}$.
Then every~$F(x)$ is a Hermitian $2 \times 2$-matrix with eigenvalues~$1$ and~$0$.
Representing~$F(x)$ as a linear combination of the identity matrix and the three
Pauli matrices~$\vec{\sigma}$ defined by
\[ \sigma^1 = \begin{pmatrix} 0 & 1 \\ 1 & 0 \end{pmatrix} , \qquad
\sigma^2 = \begin{pmatrix} 0 & -i \\ i & 0 \end{pmatrix} , \qquad
\sigma^3 = \begin{pmatrix} 1 & 0 \\ 0 & -1 \end{pmatrix} , \]
we obtain
\beq \label{Paulirep}
F(x) = \frac{1}{2} \left( \1 + \vec{v}_x \vec{\sigma} \right) \:, 
\qquad \text{where~$|\vec{v}_x|=1$} \:.
\eeq
In this way, every~$F(x)$ can be described by a unit vector~$\vec{v}_x \in S^2$. The Lagrangian
is computed to be
\beq \label{pair}
\L[A_{xy}] = \frac{1}{8}\,(1+ \vec{v}_x \!\cdot\! \vec{v}_y)^2 \:.
\eeq
In particular, all points have timelike separation. Furthermore, the Lagrangian becomes smaller if
the angle between~$\vec{v}_x$ and~$\vec{v}_y$ is larger. Thus qualitatively speaking,
our variational principle attempts to maximize the angles between the vectors~$(\vec{v}_x)_{x=1,
\ldots, m}$. In other words, the variational principle tries to distribute~$m$ points on the
sphere, maximizing their distances according to the ``repulsive pair potential''~\eqref{pair}.
This problem has been studied for a variety of potentials; see~\cite{saff+kuijlaars} for a review.

Again, the minimizer cannot be unique,
because the action of~$\U(f)$ will give rise to different minimizers.
Furthermore, other minimizers are obtained by permuting the points of~$M$.
Even if we consider the problem modulo the action of~$\U(f)$ on~$\F$ and permutations in~$M$,
in general the minimizers will still not be unique. Namely, as discussed in~\cite{saff+kuijlaars},
the discreteness of the problem will in general give rise to a complicated geometric structure
on the sphere, leading to many minima of the action. \\
\hspace*{1cm} \QEDrem \end{Example}

The following example illustrates that the causal structure can be a further
source of non-uniqueness.
\begin{Example} {{(Non-uniqueness for point distributions on~$S^2$)}}
\em We choose~$f=2$, $\beta \in (0,1)$ and let~$M=\{1,2\}$
consist of two points, again with the normalized counting measure.
Representing the matrices~$F(x)$ similar to~\eqref{Paulirep} by a linear combination of
Pauli matrices,
\beq \label{Paulirep2}
F(x) = \frac{1 - \beta}{2} \:\1 + \frac{1+\beta}{2}\: \vec{v}_x \vec{\sigma}
\qquad \text{with} \qquad |\vec{v}_x|=1 \:,
\eeq
a short calculation shows that the points~$1$ and~$2$ are spacelike separated if and only if
\[ \vec{v}_1 \!\cdot\! \vec{v}_2  < \frac{-\beta^2+6 \beta-1}{(1+\beta)^2} \:. \]
In this case, only the diagonal terms~$A_{xx}$ contribute to the action, and thus
\beq \label{Sval}
\Sact = \frac{1}{4} \left( \L[A_{11}] + \L[A_{22}] \right) = \frac{1}{4}\:(1-\beta^2)^2 \:.
\eeq
Noting that the values of~$\L[A_{xx}]$ are already determined by the given parameter~$\beta$
but are independent of the choice of the matrices~$F(x) \in \F$,
we see that~(\ref{Sval}) is even the infimum of the action.
We conclude that there is a continuous family of minimizers.
A short calculation shows that the eigenvalues of the matrix~$A_{12}$ depend on
the angle between~$\vec{v}_1$ and~$\vec{v}_1$, and thus the
minimizers for different values of this angle cannot be $\U(f)$-equivalent.
\QEDrem \end{Example}

We conclude that if the measure~$\mu$ is the counting measure, we cannot hope for
uniqueness, because the discrete nature of the problem will give rise to a complicated
geometric structure with many minima of the action.
The situation will be similarly involved if~$\mu$ consist of both discrete and a continuous parts.

However, the above examples give us hope that the uniqueness problem might simplify
if the measure~$\mu$ is purely continuous. Namely, in this case the variational principle with
Lagrangian~\eqref{pair} should have a unique minimizer, obtained by distributing the
continuous measure uniformly on the sphere. One might conjecture that the same measure should also be the minimizer in the case~$\beta>0$. Also, can one expect a unique minimizer in the case~$f \geq 2$?
Motivated by these specific questions, we now turn attention to a more systematic study of the uniqueness problem in the continuous setting.

\subsection{The Continuum Variational Principle and its Hilbert Space Formulation}
In the remainder of this chapter we shall assume that~$\mu$  is a {\em{non-atomic measure}} on~$M$
(see~\cite[Section~40]{halmosmt} or the proof Lemma~\ref{lemma14}). In this case, we can
conveniently reformulate our variational problem purely in terms of positive normalized regular
Borel measures on~$\F$, denoted in what follows by~$\B(\F)$. Namely, suppose that we
\beq \label{rhovar}
\text{minimize} \quad
\Sact(\rho) = \iint_{\F \times \F} \L[p \cdot q]\: d\rho(p) \,d\rho(q)
\quad \text{on~$\B(\F)$}\:.
\eeq
Then, since every~$F \in \M$ gives rise to a corresponding measure~$\rho$ on~$\F$
(see~(\ref{rhodef})), it is obvious that $\inf_{\F \in \M} \Sact[F] \geq \inf_{\rho \in \B(\F)} \Sact(\rho)$.
Conversely, the inductive construction in the proof of Lemma~\ref{lemma14} shows that
every~$\rho \in \B(F)$ can be realized by a function~$F \in \M$, and that this~$F$ is
unique up to isomorphisms of the measure space~$(M, \mu)$.
Hence in what follows it suffices to consider the variational principle~\eqref{rhovar},
referred to as the {\em{continuum variational principle}}. The existence of minimizers
is an immediate consequence of Lemma~\ref{lemmameasure}.
\begin{Corollary} \label{corol1}
The infimum of the continuum variational principle~\eqref{rhovar} is
attained.
\end{Corollary}

In order to get into the position to apply spectral methods, we let~$\H = L^2(\F, d\mu_\F)$
be the Hilbert space with scalar product
\[ \lbra \psi, \phi \lket = \int_\F \overline{\psi} \phi\: d\mu_\F\:, \qquad
\psi, \phi \in \H\:. \]
If we assume that the measure~$\rho$ in~\eqref{rhovar} is so regular and bounded that it
has a Radon-Nikodym decomposition (see for example~\cite[Section~31]{halmosmt})
\beq \label{RNd}
d\rho = \psi \,d\mu_\F \quad \text{with} \quad \psi \in \H\:,
\eeq
then the action can be expressed as an expectation value
\[ \Sact[\rho] = \lbra \psi, L \psi \lket\:, \]
where~$L$ is the integral operator defined by
\beq \label{Ldef}
(L \psi)(p) = \int \L[p \cdot q]\: \psi(q)\: d\mu_\F(q)\:.
\eeq
The conditions that~$\rho$ be positive and normalized can be expressed by
demanding that~$\psi \geq 0$ and that~$\lbra \psi, 1_\F \lket=1$ ($1_\F$ again denotes
the constant function one).
Hence the variational principle~(\ref{rhovar}) can be reformulated in Hilbert spaces by
\beq \label{Hvar}
\boxed{ \quad
\text{minimize} \quad
\Sact(\psi) = \langle \psi, L \psi \rangle
\quad \text{for~$\psi \in \H$ with~$\psi \geq 0$ and~$\lbra \psi, 1_\F \lket=1$}\:.
\quad }
\eeq
We remark that the constraint~$\psi \geq 0$ is
unusual in the Hilbert space setting. In particular, \eqref{Hvar} is much different from
minimizing a Rayleigh quotient. Furthermore, we point out that the
representation~(\ref{RNd}) poses a strong condition on the
Borel measure~$\rho$. However, the measures satisfying this condition are dense
in~$\B(\F)$ in the $C^0(\F)^*$-topology. Therefore, the infima of~(\ref{rhovar}) and~(\ref{Hvar})
coincide. But it is not clear whether the variational principle~(\ref{Hvar}) has a minimizer.
In cases where the answer is yes, this means that there are minimizers~$\rho$ of~(\ref{rhovar})
satisfying~(\ref{RNd}). If every minimizing sequence of~(\ref{Hvar}) had a convergent
subsequence, we could even conclude that every minimizer~$\rho$ of~(\ref{rhovar})
satisfies~(\ref{RNd}).

\subsection{Non-Uniqueness and Non-Triviality of Minimizers} \label{sec15}
In this section we shall use spectral methods to prove the following result.
\begin{Thm} \label{thm2}
For the continuum variational principle in the Hilbert space formulation~\eqref{Hvar}
the following holds:
\begin{itemize}
\item[(i)] In the case~$\beta=0$, the constant function~$1_\F$ is a minimizer,
\[ \lbra 1_\F, L 1_\F \lket = \inf_{\rho \in \B(\F)} \Sact(\rho)\:. \]
There is an infinite-dimensional family of minimizers.
\item[(ii)] In the case~$0<\beta<1$, the function~$1_\F$ is not a minimizer,
\[ \lbra 1_\F, L 1_\F \lket > \inf_{\rho \in \B(\F)} \Sact(\rho)\:. \]
The minimizer~$\rho \in \B(\F)$ from Corollary~\ref{corol1} is not unique.
\end{itemize}
\end{Thm} \noindent
Moreover, in the case~$\beta=0$ and~$f=2$, we give an explicit example of a
minimizer~$\rho \in \B(\F)$ which does not have the representation~\eqref{RNd}
with~$\psi \in \H$ (see Example~\ref{ex3}). 

In preparation for the proof, we compile a few spectral properties of the operator~$L$.
\begin{Lemma} $L$ is a compact self-adjoint operator on~$\H$. The eigenvectors corresponding
to non-zero eigenvalues are continuous functions on~$\F$.
The sup-norm $\|L\|$ is a non-degenerate
eigenvalue, and the constant function~$1_\F$ is the corresponding
eigenvector.
\end{Lemma} 
\Proof Clearly, the kernel~$\L[p \,q]$ of~$L$ is a continuous function on~$\F \times \F$.
Using furthermore that~$\F$ is a compact manifold, the operator~$L$ is obviously bounded.
Moreover, the kernel~$\L[p \cdot q]$ is real-valued and, as explained after Definition~\ref{def1}, it is symmetric in its two arguments. This implies that~$L$ is self-adjoint.
The integral estimate
\[ \Tr(L^* L) = \iint_{\F \times \F} \L[p \, q] \:\L[q \, p] \:d\mu_\F(p)\: d\mu_\F(q)
\:\leq\: \mu(\F)^2 \sup_{p,q}|\L[p \, q]|^2 \]
shows that the operator~$L$ is Hilbert-Schmidt, and thus compact.
This means that its the spectrum~$\sigma(L) \subset \R$ is purely discrete, bounded and accumulates
at most at zero. The spectral theorem gives a decomposition
\[ L = \sum_{\lambda \in \sigma(L)} \lambda E_\lambda\:, \]
where the~$E_\lambda$ are projectors onto finite-dimensional, mutually orthogonal eigenspaces.

Suppose that~$\psi$ is an eigenvector of~$L$ corresponding to an eigenvalue~$\lambda \neq 0$.
Then~$\psi$ can be written as
\[ \psi(p) = \frac{1}{\lambda}\: \int_{\F} \L[p \,q] \:\psi(q)\: d\mu_\F(q)\:. \]
Regarding the right side as a convolution of~$\psi$ with a continuous kernel, one sees
that~$\psi$ is a continuous function.

The Lagrangian~\eqref{LSdef} is $\U(f)$-invariant, meaning that
$\L[p\, q] = \L[Up\, qU^{-1}]$ for every~$U \in \U(f)$.
Using furthermore the $\U(f)$-invariance of the integration measure~$d\mu_\F$, we obtain
\[ (L \,1_\F)(p) = \int_{\F} \L[p\, q] \:d\mu_\F(q) 
= \int_{\F} \L[p\, U^{-1} q U] \:d\mu_\F(q)
= (L \,1_\F)(U p\, U^{-1}) \:. \]
Since~$\U(f)$ acts transitively on~$\F$, we conclude that the function $L(1_\F)$ is constant.
Hence~$1_\F$ is an eigenvector.

The idea for completing the proof is to note that the operator~$L$ has a non-negative kernel
(cf.~\eqref{Ldef} and~\eqref{LSdef}) and to apply the Perron-Frobenius theorem
(see~\cite[Chapter~5]{serreM} for matrices and~\cite[Section~3.3]{glimm+jaffe} for integral
operators). Unfortunately, this theorem cannot be applied in our setting, because it requires
that the kernel be positive almost everywhere,
whereas our kernel~$L[p\,q]$ vanishes whenever the points~$p$ and~$q$ are
spacelike separated. But we can adapt the Perron-Frobenius method as follows:
First, the estimate
\[ -\inf \sigma(L) = \sup_{\phi \in \H,\; \|\phi\|=1} -\lbra \phi, L \phi \lket 
\leq \sup_{\phi \in \H,\; \|\phi\|=1} \lbra |\phi|, L |\phi| \lket \;\leq\; \sup \sigma(L) \]
shows that the supremum of the spectrum coincides with the sup-norm.
Hence, using that~$L$ is compact, $\|L\|>0$
is an eigenvalue corresponding to a finite-dimensional eigenspace.
Suppose that~$\psi$ is a corresponding normalized eigenvector. Then
\[ \|L\| = \lbra \psi, L \psi \lket
\leq \lbra |\psi|, L |\psi |\lket \leq \|L\| \:, \]
showing that~$|\psi|$ is again an eigenvector corresponding to the eigenvalue~$\|L\|$.
This eigenvector is not orthogonal to $1_\F$, because
\[ \lbra |\phi|, 1_\F \lket = \int_\F |\phi|\: d\mu_\F > 0 \:. \]
This implies that the two eigenvectors must be in the same eigenspace, and
thus $L \,1_\F = \|L\|\, 1_\F$.

We now proceed by contradiction. Assume that the eigenvalue~$\|L\|$ is degenerate.
We choose a vector~$\psi$ in the corresponding eigenspace which is orthogonal to~$1_\F$.
Since the real and imaginary parts of an eigenfunction are again eigenvectors,
we can arrange that $\psi$ is real-valued. Then the orthogonality $\lbra \psi, 1_\F \lket=0$ implies
that~$\psi$ 
changes sign. Since~$\psi$ is continuous, its zero set must be non-empty. We choose a point~$p_0$ on
the boundary of the zero set. Then $\psi(p_0)=0$, but $\psi$ is non-trivial on any neighborhood
of~$p_0$. Since the Lagrangian is continuous and $\L[p_0 \, p_0] > 0$, there is a point~$q_0$ in a neighborhood of~$p_0$ where~$\L[p_0 \, q_0] > 0$ and~$\psi(q_0) \neq 0$. 
Taking the linear combination
\[ \phi = 1_\F - \frac{2}{\psi(q_0)}\: \psi \:, \]
we have constructed an eigenvector corresponding to the eigenvalue~$\|L\|$ such that
\[ \phi(p_0)=1\:,\quad \phi(q_0) =-1 \qquad \text{and} \qquad \L[p_0\, q_0] > 0\:. \]
Using continuity, it follows that the function $\L[p\, q]\, \phi(p)\, \phi(q)$ is strictly negative on a set of
non-zero measure. Hence
\[ \begin{split}
\|L\|\:\|\phi\|^2 = \lbra \phi, L \phi \lket &= \iint_{\F \times \F}
\L[p\, q]\, \phi(p)\, \phi(q)\: d\mu_\F(p)\: d\mu_\F(q) \\
&< \iint_{\F \times \F}
\L[p\, q]\, |\phi(p)|\, |\phi(q)|\: d\mu_\F(p)\: d\mu_\F(q) =  \lbra |\phi|, L |\phi| \lket \:, 
\end{split} \]
a contradiction.
\QED

We can now characterize the structure of the minimizers of the continuum variational principle
depending on spectral properties of~$L$. We distinguish between three cases:
\begin{itemize}
\item[(A)] $L \geq 0$ and $\ker L = \{0\}$: \\
We choose an orthonormal eigenvector basis~$(\psi_n)_{n \in \N}$ with~$L \psi_n
= \lambda_n \psi_n$, ordered such that~$\psi_1=1_F$ and~$\lambda_1=\|L\|$.
Expanding the vector~$\psi$ in~\eqref{Hvar}
in this basis, the condition~$\lbra \psi, 1_\F \lket=1$ implies that the coefficient of~$\psi_1$
equals one,
\beq \label{psirep}
\psi = \psi_1 + \sum_{n=2}^\infty c_n \psi_n \qquad \text{with} \qquad
\sum_{n=2}^\infty |c_n|^2 < \infty\:.
\eeq
Using furthermore that all eigenvalues are strictly positive, we obtain the estimate
\beq \label{estim}
\lbra \psi, L \psi \lket = \lambda_1 + \sum_{n=2}^\infty \lambda_n |c_n|^2 
\geq \lambda_1\:,
\eeq
and equality holds only if all the coefficients~$c_2, c_3, \ldots$ vanish.
This shows that the constant function is the unique minimizer.
\item[(B)] $L \geq 0$ and the kernel of~$L$ is non-trivial: \\
Again representing~$\psi$ in the form~\eqref{psirep}, the inequality~\eqref{estim} again
holds, but we have equality if and only if $\psi-\psi_1 \in \ker L$. Since we must take into
account the condition~$\psi \geq 0$, the set of all minimizers in~$\H$ is given by
\beq \label{minfamily}
\left( \psi_1 + \ker L \right) \cap \{ \psi \in \H, \psi \geq 0 {\text{ a.e.}} \}\:.
\eeq
Thus the constant function is again a minimizer, but there are other, nontrivial minimizers.
If the kernel of~$L$ is finite-dimensional, the functions in~\eqref{minfamily} are all
continuous, and thus we can say that all minimizing measures of the variational principle~\eqref{rhovar}
have the representation~\eqref{RNd} with a continuous function~$\psi \in \H$.
However, if~$L$ is infinite-dimensional, in general there will be
sequences in~\eqref{minfamily} which do not converge in~$\H$ but do converge in~$C^0(\F)^*$
to a non-trivial Borel measure.
Thus in this case we can expect minimizers of~\eqref{rhovar} which do not admit a
representation~\eqref{RNd} with~$\psi \in \H$
(for an example of such a minimizer see Example~\ref{ex3}).
\item[(C)] $L$ has negative eigenvalues: \label{caseC} \\ 
The inequality~\eqref{estim} can be violated by choosing the coefficient~$c_n$ corresponding
to one of the negative eigenvalues to be non-zero. Hence the constant function is no longer
a minimizer. There seems no general reason why a minimizing sequence should converge in~$\H$.
Thus we cannot expect that there are any minimizers in~$\H$. But according to
Corollary~\ref{corol1}, the minimizers will still exist in the sense of measures. Since these
minimizers necessarily break the $\U(f)$-symmetry,
the action of~$\U(f)$  (see page~\pageref{transitive}) will give rise to other minimizers.
Hence the minimizer~$\rho \in \B(\F)$ as obtained from Corollary~\ref{corol1} is not unique.
\end{itemize}
Following these arguments, it remains to analyze the spectrum of~$L$.
We begin with the case~$\beta=0$.
\begin{Lemma} \label{betazero}
In the case~$\beta=0$, the operator~$L$ is non-negative. Its rank is finite and bounded by
\[ \dim L(\H) < f^4\:. \]
\end{Lemma}
\Proof In the case~$\beta=0$, every~$p \in \F$ is a $f \times f$-matrix of rank one, having the
non-trivial eigenvalue one.
We denote a normalized eigenvector of~$p$ corresponding to the non-trivial eigenvalue
by~$u(p) \in \C^f$. Since the matrix~$p\, q$ has rank at most one, the Lagrangian~\eqref{LSdef}
simplifies to
\beq \label{Lbetaz}
\L[p \, q] = \frac{1}{2} \left| \lbra u(p), u(q)\lket_{\C^f} \right|^4\:.
\eeq
We now introduce the operator~$K\,:\, \H \rightarrow \C^{f} \otimes \C^{f} \otimes (\C^{f})^* \otimes
(\C^{f})^*$ by
\[ K(\psi) = \int_{\F} \psi(p)\: u(p) \otimes u(p) \otimes u(p)^* \otimes u(p)^*\: d\mu_\F(p) \]
(where we use the natural embedding of $\C^f$ to its dual space given by complex conjugation).
Then
\begin{align*}
\|K(\psi)\|^2 &= \iint_{\F \times \F} \overline{\psi(p)} \psi(q)\:
\lbra u(p), u(q) \lket_{\C^f}^2 \:\overline{\lbra u(p), u(q) \lket_{\C^f}^2}\:
d\mu_\F(p)\, d\mu_\F(q) \\
&\!\!\!\overset{\eqref{Lbetaz}}{=} 2 \lbra \psi, L \psi \lket_{\H}\,.
\end{align*}
In other words, $2 L=K^* K$, showing that~$L$ is non-negative and that its rank is at most
the dimension of the vector space~$\C^{f} \otimes \C^{f} \otimes (\C^{f})^* \otimes
(\C^{f})^*$.
\QED
We next give a more explicit analysis of the case~$f=2$ and~$\beta=0$. 
In this case, the Pauli representation~\eqref{Paulirep} allows us to identify~$\F$ with~$S^2$.
Then~$L$ is a spherically symmetric operator on~$L^2(S^2)$. Thus it has the same
eigenspaces as the spherical Laplacian,
\[ L = \sum_{l=0}^\infty \lambda_l \:E_l \:, \]
where the~$E_l$ are the projection operators onto the eigenspaces of the spherical Laplacian
corresponding to the eigenvalue~$l(l+1)$. The kernel of the~$E_l$ can be given in terms of
the spherical harmonics~$Y^m_l$ by
\[ E_l(p,q) = 4 \pi \sum_{m=-l}^l Y^m_l(p)\: \overline{Y^m_l(q)} ,\quad 
\text{where $p,q \in S^2$} \]
(the factor~$4 \pi$ arises because our integration measure~$\mu_\F$ on~$S^2$ has total volume one).
The eigenvalues~$\lambda_l$ are most easily computed by applying~$L$ to the
spherical harmonic~$Y^0_l$,
\beq \label{LY}
\lambda_l Y^0_l(p) = (L \,Y^0_l)(p) = \int_\F \L[p\, q]\: Y^0_l(q)\: d\mu_\F(q) \:.
\eeq
Choosing on~$S^2$ standard polar coordinates~$(\vartheta, \varphi)$, 
the spherical harmonics~$Y^0_l$ are multiples of the Legendre polynomials~$P_l(\cos \vartheta)$.
More precisely, using the standard normalization conventions
\[ Y^0_l(\vartheta, \varphi) = \sqrt{\frac{2l+1}{4 \pi}}\: P_l(\cos \vartheta)\qquad \text{and} \qquad
P_l(1)=1 \:, \]
we can evaluate~(\ref{LY}) for $p$ at the north pole~$\vartheta=0$ to obtain the simple formula
\beq \label{lcomp}
\lambda_l = \frac{1}{2} 
\int_{-1}^1 \L(\cos \vartheta)\:P_l(\cos \vartheta)\: d\cos \vartheta\:.
\eeq
Setting~$\beta=0$, the Lagrangian simplifies to~\eqref{pair}, 
\[ \L(\cos \vartheta) = \frac{1}{8}\, (1+\cos \vartheta)^2  \qquad
\text{(if $f=2$, $\beta=0$)}\:. \]
In this case, the integral~\eqref{lcomp} can easily be calculated 
\[ \lambda_0 = \frac{1}{6}\:,\quad \lambda_1 = \frac{1}{12}, \quad
\lambda_2 = \frac{1}{60}\:,\qquad \lambda_3 = \lambda_4=\ldots=0\:.\]
Counting the multiplicities $2l+1$ of the eigenspaces, the rank of~$L$ is computed to
be~$1+3+5=9$, in agreement with the upper bound~$f^4=16$ from Lemma~\ref{betazero}.

This detailed information allows us to give a minimizer which is not in~$\H$.
\begin{Example} (A distributional minimizer)
\label{ex3} \em In the case~$f=2$ and~$\beta=0$, we again use the
identification~$\F \simeq S^2$ and choose polar coordinates~$(\vartheta, \varphi$).
For any parameter~$a \in [0,1]$, we define the measure~$\rho \in \B(\F)$ by
\[ \begin{split}
\int_\F g \:d\rho = & \int_\F \left[ a + \frac{1-a}{2}\: \Theta \Big(|\cos \vartheta|-\frac{1}{2} \Big) \right]
g\: d\mu_\F \\
&+ \frac{3}{8}\: (1-a)\:\frac{1}{2 \pi} \int_0^{2 \pi}
\left[ g \Big(\cos \vartheta=\frac{1}{2}, \varphi \Big) + g \Big(
\cos \vartheta=-\frac{1}{2}, \varphi \Big) \right] d\varphi\:,
\end{split} \]
where~$\Theta$ is the Heaviside function.
This is a positive normalized regular Borel measure, but due to the singular contributions
at $\cos \vartheta = \pm \frac{1}{2}$, it cannot be represented in the form~\eqref{RNd}
with~$\psi \in \H$. In order to show that~$\rho$ is a minimizer, we must verify that it coincides
with the constant function on the nontrivial eigenspaces of~$L$, i.e.
\[ \rho(Y^0_0) = \frac{1}{\sqrt{4 \pi}} \qquad \text{and} \qquad
\rho(Y^m_l) = 0 \quad \text{for all } l=1,2 \text{ and } m=-l,\ldots, l\:. \]
The first condition is obvious because $Y^0_0=(4 \pi)^{-\frac{1}{2}}$ is constant and~$\rho$
is normalized. Since the spherical harmonics~$Y^m_l$ are the restrictions to~$S^2$ of polynomials
of degree~$l$ in~$\R^3$, a symmetry consideration in the $\varphi$-integral shows that all the
other conditions reduce to the three constraints
\beq \label{cc2}
\rho(z) = 0 \:,\qquad
\rho (3 z^2-1)=0 \:,\qquad
\rho (3 x^2-1)=0\:,
\eeq
where~$(x,y,z)$ are Cartesian coordinates, where~$\vartheta$ denotes the angle to the
$z$-axis. The first equation in~\eqref{cc2} is immediate by symmetry,
whereas the second equation follows from the computation
\[ \frac{1-a}{4} \int_{\frac{1}{2}}^1 (3z^2-1)\:dz \:+\: \frac{3}{8}\:(1-a)\: (3z^2-1)
\Big|_{z=\frac{1}{2}}
= 0 \:. \]
In order to verify the third relation, we first note that the symmetry around the $z$-axis
yields~$\rho(3x^2-1)=\rho(3y^2-1)$. Also using the last equation in~\eqref{cc2}, we obtain
\[ \rho(3x^2-1) = \frac{1}{2} \:\rho \Big(3(x^2+y^2+z^2)-3 \Big) = \frac{1}{2}\:\rho(3-3) = 0\:, \]
concluding the proof. \QEDrem
\end{Example}
\vspace*{0.5em}

\begin{figure}
\begin{center}
\includegraphics[width=6.5cm]{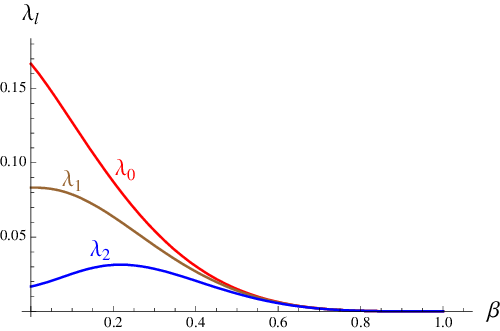} $\quad$
\includegraphics[width=6.5cm]{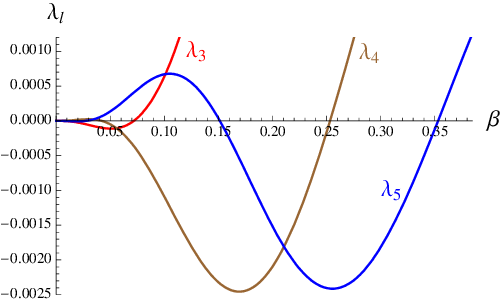}
\caption{The eigenvalues~$\lambda_l(\beta)$ of the operator~$L$ in the case~$f=2$.}
\label{fig1}
\end{center}
\end{figure}

If~$\beta>0$ (and still~$f=2$), the situation is more interesting because of the non-trivial
causal structure, which leads to a region where the Lagrangian vanishes identically.
Namely, a short calculation using~\eqref{Paulirep2} yields
\beq \label{Lft}
\L(\cos \vartheta) = \frac{(1+\beta)^4}{8}\: (1+\cos \vartheta)
\: \max \!\left( 0, \;
\cos \vartheta + \frac{1-6\beta+\beta^2}{(1+\beta)^2} \right) .
\eeq
The resulting integrand in~\eqref{lcomp} is again a polynomial in $\cos \vartheta$,
and thus for every~$l$, the function~$\lambda_l(\beta)$ can be computed in closed form.
For example,
\[ \lambda_0 = \frac{(1-\beta)^4 (1+4 \beta + \beta^2)}{6 (1+\beta)^2}\:, \]
and similarly for the other eigenvalues (clearly, the formulas get more complicated for larger~$l$,
but expressions up to $l \approx 20$ are handled easily by computer algebra).
On the left of Figure~\ref{fig1} the three lowest eigenvalues~$\lambda_0$, $\lambda_1$
and~$\lambda_2$ of~$L$ are shown as functions of~$\beta$.
These eigenvalues are always positive. On the right of Figure~\ref{fig1}, the next
eigenvalues~$\lambda_3$, $\lambda_4$ and~$\lambda_5$ are plotted versus~$\beta$. The
Taylor expansion
\[ \lambda_3 = -\frac{16}{3}\: \beta^3 + \O(\beta^4) \]
shows that~$\lambda_3$ is negative for small positive~$\beta$, and Figure~\ref{fig1}
illustrates that this function~$\lambda_3(\beta)$ stays negative on the interval $0<\beta<0.07$.
Thus for~$\beta$ in this range, we are in case~(C) on page~\pageref{caseC} where the minimizer is
non-trivial. If~$\beta$ is further increased,
$\lambda_3$ becomes positive, but then~$\lambda_4$ is negative. If~$\beta$ is further increased,
$\lambda_4$ becomes positive, but then~$\lambda_5$ is negative.
More generally, these plots suggest that for every~$\beta$ in the range~$0<\beta<1$ at least
one of the~$\lambda_l$ should be negative. This is indeed the case, as the following lemma shows.
\begin{Lemma} \label{lemma112} In the case~$f=2$,
\[ \min_{l \in \N_0} \lambda_l(\beta) < 0 \qquad \text{for all~$\beta \in (0,1)$}\:. \]
\end{Lemma}
\Proof In order to analyze the asymptotics $\beta \nearrow 1$, we first expand the second
fraction in~\eqref{Lft},
\[ \frac{1-6\beta+\beta^2}{(1+\beta)^2} = -1 + \frac{1}{2}\: (1-\beta)^2 + \O((1-\beta)^3)\:, \]
showing that the Lagrangian vanishes identically except in a neighborhood of the north
pole~$\vartheta=0$. Thus we may also expand in powers of~$\vartheta$ to obtain the
asymptotic Lagrangian
\beq \label{Lasy}
\L^\text{asy}(\vartheta) = \frac{(1+\beta)^4}{8}\: \left(1-\frac{\vartheta^2}{4} \right)
\: \max \!\left( 0, \; \frac{4 (1-\beta)^2}{(1+\beta)^2}-\vartheta^2 \right) .
\eeq
In order to derive the asymptotic form of the Legendre polynomials~$P_l(\cos \vartheta)$ near
the pole, we note that these polynomials are given as solutions of the ODE
\[ -\frac{1}{\sin \vartheta} \frac{d}{d\vartheta} \left( \sin \vartheta\: \frac{d}{d\vartheta}
P_l(\cos \vartheta) \right) = l(l+1)\:P_l(\cos \vartheta) \:,\qquad P_l(1)=1\:. \] 
Using the asymptotics~$\sin \vartheta = \vartheta + \O(\vartheta^3)$, this differential equation simplifies to
\[ -\left( \frac{d^2}{d\vartheta^2} + \frac{1}{\vartheta} \frac{d}{d\vartheta} \right) P^\text{asy}_l(\vartheta) = l(l+1)\:P^\text{asy}_l(\vartheta) \:,\qquad P^\text{asy}_l(0)=1\:, \]
whose solution is a Bessel function of the first kind,
\beq \label{Pasy}
P^\text{asy}_l(\vartheta) = J_0 \!\left(\sqrt{l(l+1)}\: \vartheta \right)\:.
\eeq
Substituting~\eqref{Lasy} and~\eqref{Pasy} into~\eqref{lcomp} and using the asymptotic form
of the integration measure $d\cos \vartheta = (\vartheta+ \O(\vartheta^3))\, d\vartheta$,
the integral can be computed in terms of the confluent hypergeometric function,
\begin{align}
\lambda^\text{asy}_l &= \frac{1}{2} \int_0^\pi \L^\text{asy}(\vartheta)\: P^\text{asy}_l(\vartheta)\: 
\vartheta\, d\vartheta \label{lasy} \\
&= \frac{(1-\beta)^2}{\lambda} \Big(
(1 + \beta (2 + \beta + \lambda)) \;{_0}F_1(3,-x) - (1-\beta)^2 \;{_0}F_1(2,-x) \Big)\:, \label{lexp}
\end{align}
where we set
\beq \label{xdef}
\lambda = l(l+1) \quad \text{and} \quad x = \frac{(1-\beta)^2}{(1+\beta)^2}\:\lambda\:.
\eeq

Next we need to specify~$l$ as a function of~$\beta$.
Qualitatively, speaking, the Bessel function~\eqref{Pasy} oscillates with frequency
of the order~$\sqrt{l(l+1)}$, whereas the Lagrangian~\eqref{Lasy} decreases in~$\vartheta$,
vanishing identically for
\[ \vartheta \geq \vartheta_\text{max} := 2 \:\frac{1-\beta}{1+\beta}\:. \]
Thus in order to make the integral~\eqref{lasy} negative, we choose~$l$ such
that~$P^\text{asy}_l(\vartheta)$ oscillates on~$[0, \vartheta_\text{max}]$ just once, being negative
at~$\vartheta=\vartheta_\text{max}$. A good method would be to determine~$l$ from the equation
\[ \sqrt{l_\text{asy}(l_\text{asy}+1)} \;\vartheta_\text{max} = 5.5 \:. \]
Since this is in general not an integer, we introduce~$l$ using the Gauss bracket,
\[ l(\beta) := 1 + \left[ l_\text{asy} \right]\:. \]
With this choice of~$l$, a simple calculation shows that if~$\beta \in [0.4,1)$, the 
variable~$x$ as defined by~\eqref{xdef} takes values in the range~$x \in [7, 12]$.
For~$\beta$ and~$x$ chosen in these intervals, a direct inspection shows that
the function~\eqref{lexp} is always negative.

\begin{figure}
\begin{center}
\includegraphics[width=8.2cm]{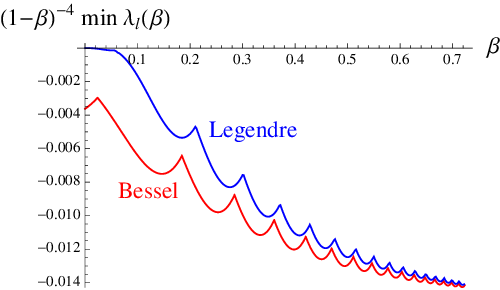}
\caption{The smallest eigenvalue of~$L$ in the case~$f=2$
using the exact formulas with Legendre polynomials and the asymptotics with Bessel functions.}
\label{fig2}
\end{center}
\end{figure}

Our analysis so far shows that the lemma holds if~$\beta$ is sufficiently close to one.
On the other hand, on any interval~$\beta \in (0, \beta_\text{max}]$ with~$\beta_\text{max}<1$,
one can consider the explicit formulas obtained from the Legendre polynomials to verify
that~$L$ has a negative eigenvalue. 
As shown in Figure~\ref{fig2}, at~$\beta \approx 0.7$ the asymptotics with Bessel functions is a
good approximation to the exact analysis.
Because of this obvious fit, we omit rigorous error estimates of the asymptotic analysis.
\QED

The case~$f>2$ is considerably more complicated.
However, as shown in the next lemma, there is a general mechanism giving rise to negative
eigenvalues.

\begin{Lemma} \label{lemma113}
If~$f>2$ and~$0<\beta<1$, the operator~$L$ has negative eigenvalues.
\end{Lemma}
\Proof We choose~$p, q \in \F$ having the matrix representation
\[ p = P_{e_1} - \beta P_{e_2} \:,\qquad q = P_{e_1} - \beta P_{e_3}\:, \]
where~$P_{e_i}$ denotes the orthogonal projection to the canonical basis vector~$e_i \in \C^f$.
It suffices to show that the $2 \times 2$-matrix
\beq \label{lm}
l := \begin{pmatrix} \L[p\,p] & \L[p\,q] \\ \L[q\,p] & \L[q\,q] \end{pmatrix}
\eeq
has a negative eigenvalue, as the following argument shows. Suppose that~$\phi \in \C^2$ is a vector
with~$\lbra \phi, l \phi \lket_{\C^2}<0$. We consider a series~$\psi_n \in \H$ which converges
in the $C^0(\F)^*$-topology to the measure $\phi_1 \delta_p + \phi_2 \delta_q$
(where~$\delta_p$ denotes the Dirac measure supported at~$p$). Then
\[ \lbra \psi_n, L \psi_n \lket_\H \:\xrightarrow{n \rightarrow \infty}\:
\lbra \phi, l \phi \lket_{\C^2}<0 \:, \]
proving that~$L$ has indeed negative eigenvalues.

To show that the matrix~\eqref{lm} has a negative eigenvalue, we first compute the spectra of
the matrix products,
\[ \sigma(p \, p) = \sigma(q\, q) = \{1, \beta^2, 0\} \:,\quad
 \sigma(p \, q) = \sigma(q\, p) = \{1, 0\}\:. \]
Hence
\[ l = \frac{1}{2} \begin{pmatrix} (1-\beta^2)^2 & 1 \\  1 & (1-\beta^2)^2 \end{pmatrix} 
\:, \qquad
\det l = \frac{1}{4} \left( (1-\beta^2)^4 - 1 \right) < 0\:,  \]
and thus~$l$ has precisely one negative eigenvalue.
\QED
This completes the proof of Theorem~\ref{thm2}. 

We now turn attention to a more general class of variational principles.
The main additional difficulty will be that the target space is no longer compact.
Intuitively, this leads to the possibility that part of the support
of the measures~$\rho_k$ ``escapes to infinity.'' In order to rule this out, we must
control the behavior of the measures~$\rho_k$ by suitable a-priori estimates.

\section{A General Class of Causal Variational Principles} \label{sec2}
\subsection{Definitions and Statement of Results} \label{sec21}
Let~$(M, \mu)$ be a measure space of total volume~$\mu(M)=1$.
For given integers~$f, n$ with~$f \geq 2n$ we let~$\F$ be the set of all Hermitian
$f \times f$-matrices of rank at most~$2n$, having at most~$n$ positive and at most~$n$
negative eigenvalues. We let~$\M$ be the set of matrix-valued functions
\[ \M = \left\{ F : M \rightarrow \F \text{ measurable} \right\} . \]
For a given~$F \in \M$ and any~$x, y \in M$, we again form the matrix product~\eqref{Adef}
and denote its eigenvalues counted with algebraic multiplicities by
\beq \label{lambdacount}
\lambda_1^{xy}, \ldots, \lambda_{2n}^{xy}, 
\underbrace{0,\ldots, 0}_{\text{$f-2n$ times}}
\quad \text{with} \quad \lambda_j^{xy} \in \C \:.
\eeq
We define the {\em{spectral weight}}~$|A_{xy}|$ by
\beq \label{sweight1}
|A_{xy}| = \sum_{j=1}^{2n} |\lambda_j^{xy}| ,
\eeq
and similarly set~$|A_{xy}^2| = \sum_{j=1}^{2n} |\lambda_j^{xy}|^2$.
We introduce
\beq \label{Lagdef}
\boxed{ \quad \text{the Lagrangian} \quad\quad \L[A_{xy}] = |A_{xy}^2| - \frac{1}{2n}\: |A_{xy}|^2 
\quad}
\eeq
and define the functionals~$\Sact$ and~$\T$ by
\begin{align}
\Sact[F] &= \iint_{M \times M} \L[A_{xy}]\: d\mu(x)\, d\mu(y) \label{Sdef} \\
\T[F] &= \iint_{M \times M} |A_{xy}|^2\: d\mu(x)\, d\mu(y) \label{Tdef}\:.
\end{align}
We also introduce the following constraints:
\begin{itemize}
\item[(C1)] The {\bf{trace constraint}}: \label{pagetc0}
\[ \int_M \Tr (F(x)) \, d\mu(x) = f\:. \]
\item[(C2)] The {\bf{identity constraint}}:
\[ \int_M F(x) \, d\mu(x) = \1_{\C^f}\:. \]
\item[(C3)] {\bf{Prescribing $2n$ eigenvalues}}: We denote the eigenvalues of~$F(x)$ counted with
multiplicities by~$\nu_1, \ldots, \nu_f$ and order them such that
\beq \label{nudenote}
\nu_1 \leq \cdots \leq \nu_{n} \;\leq 0 \;\leq\; \nu_{n+1} \leq \ldots \leq \nu_{2n}
\eeq
and~$\nu_{2n+1}=\cdots=\nu_f=0$. We introduce constants~$c_1,\ldots, c_{2n}$ and impose that
\beq \label{pin}
\nu_j(x) = c_j \qquad \text{for all } x \in M \text{ and } j \in \{1, \ldots, 2n\} \:.
\eeq
\end{itemize}
We now state our results and explain them afterwards.
In order to rule out trivial cases, we shall always assume that the set of functions satisfying
the constraints is non-empty.
\begin{Thm} \label{thm3} Imposing the constraint~(C3) and in addition possibly
the constraints (C1) or~(C2), the variational principle
\[ \text{minimize $\Sact[F]$ on~$\M$} \]
attains its minimum.
\end{Thm}

\begin{Thm} \label{thm4} For any parameter~$\nu$ with
\beq \label{nucond}
\nu > - \frac{2n}{2n-1}\:,
\eeq
we consider the variational principle
\beq \label{TSmin}
\text{minimize $\T[F] + \nu\, \Sact[F]$ on~$\M$} \:,
\eeq
possibly with the additional constraints~(C1) or~(C2). Then the minimum is attained
by a function~$F \in L^2(M, \F, d\mu)$.
\end{Thm}

\begin{Thm} \label{thm5} For any parameter~$C > 0$, we consider the variational principle
\[ \text{minimize $\Sact[F]$~on~$\M_C := \{F \in \M$ with $\T[F] \leq C \}$} \:, \]
possibly with the additional constraints~(C1) or~(C2). Then the minimum is attained
by a function~$F \in L^2(M, \F, d\mu)$.
\end{Thm}

Before coming to the proofs, we briefly discuss the variational principles and our results.
First of all, labeling the eigenvalues as in~\eqref{lambdacount}, the Lagrangian~\eqref{Lagdef}
can be written as
\beq \label{Lcrit}
\L[A_{xy}] = \frac{1}{4n} \sum_{i,j=1}^{2n} \Big(|\lambda^{xy}_i| - |\lambda^{xy}_j| \Big)^2 .
\eeq
In particular, one sees that the Lagrangian is always non-negative.
Furthermore, the Lagrangian is causal in the sense that it vanishes identically for spacelike
separation defined as follows.
\begin{Def} {\bf{(causal structure)}} \label{causal}
Two space-time points~$x,y \in M$ are called {\bf{timelike}}
separated if the~$\lambda^{xy}_j$ are all real. They are said to be
{\bf{spacelike}} separated if all the~$\lambda^{xy}_j$, $j=1,\ldots, 2n$ form complex
conjugate pairs and all have the same absolute value.
In all other cases, the points~$x$ and~$y$ are said to be {\bf{lightlike}} separated.
\end{Def}

Theorem~\ref{thm3} is an obvious generalization of Theorem~\ref{thm1}.
Theorems~\ref{thm4} and~\ref{thm5} are more interesting because the range of the admissible
functions~$F$ is in general a non-compact subset of~$\F$. The constraints~(C1), (C2)
and/or (C3) are needed in order to rule out the trivial minimizer~$F \equiv 0$.
Clearly, (C2) implies~(C1). To explain the assumptions
of Theorem~\ref{thm4}, we note that the inequality
\[ |A_{xy}^2| = \sum_{j=1}^{2n} |\lambda_j^{xy}|^2 \leq \left( \sum_{j=1}^{2n} |\lambda_j^{xy}|
\right)^2 = |A_{xy}|^2 \]
yields the following upper bound for the Lagrangian~\eqref{Lagdef},
\beq \label{Les}
\L[A_{xy}] \leq \left(1-\frac{1}{2n} \right)\: |A_{xy}|^2\:.
\eeq
Thus the condition~\eqref{nucond} ensures that the functional~$\T[F] + \nu\, \Sact[F]$
in Theorem~\ref{thm4} is non-negative.
If~$\nu < -\frac{2n}{2n-1}$, this functional is unbounded from below, as the following
example shows.
\begin{Example} (Ill-posedness)
\em We consider the case~$n=1$ and~$f=2$, and choose~$M=\{1,\dots,4\}$
with the normalized counting measure~$\mu(\{l\})=\frac{1}{4}$ for all~$l \in M$.
For any~$k \geq 0$ we define~$F \in \M$ by
\begin{align*}
F(1) &= \begin{pmatrix} k+4 & 0 \\ 0 & 0 \end{pmatrix} ,&
F(2) &= \begin{pmatrix} 0 & 0 \\ 0 & k+4 \end{pmatrix} , \\
F(3) &= \begin{pmatrix} -k & 0 \\ 0 & 0 \end{pmatrix} ,&
F(4) &= \begin{pmatrix} 0 & 0 \\ 0 & -k \end{pmatrix} .
\end{align*}
Then the identity constraint~(C2) is satisfied. Moreover, since the matrices~$A_{xy}$
are all of rank at most one, we know that~$|A_{xy}^2| = |A_{xy}|^2$. Thus
\[ \T[F] + \nu\, \Sact[F] = \frac{1}{16}
\left(1+\nu \: \frac{2n-1}{2n} \right) \sum_{x,y \in M} |A_{xy}|^2 \:. \]
In the case~$\nu<-\frac{2n}{2n-1}$, the bracket is negative, and thus the functional tends
to minus infinity as~$k \rightarrow \infty$. Hence the variational principle~\eqref{TSmin}
is ill-posed.
\QEDrem
\end{Example} \noindent
The remaining border case~$\nu=-\frac{2n}{2n-1}$ will not be considered in this paper.

To explain Theorem~\ref{thm5}, we first note that according to~\eqref{Lcrit}, the action~$\Sact$
in Theorem~\ref{thm5} is non-negative.
As will be explained in the next Section~\ref{sec22}, the constraint~$\T[F] \leq C$ is needed,
because without an a-priori bound on~$\T$ there are examples of divergent
minimizing sequences.
We also point out that the we cannot prove Theorem~\ref{thm5} with the
alternative constraint~$T[F]=C$. This is because when taking the limit of a minimizing
sequence~$F_k \rightarrow F$, the functional~$\T$ need not converge, and indeed
we can only prove that~$\T[F] < \lim_{k \rightarrow \infty} \T[F_k]$. This so-called
``bubbling phenomenon'' will be explained in Section~\ref{sec23}.
The proofs of Theorems~\ref{thm3}--\ref{thm5} will be completed in
Section~\ref{sec24}.

\subsection{Counter Examples to Compactness} \label{sec22}
We now illustrate in simple counter examples why the constraint~$\T[F] \leq C$
in Theorem~\ref{thm5} is needed. In the first three examples, the action~$\Sact[F]$ is
uniformly bounded for a divergent family of functions~$F$.
In the last example, we construct an unbounded series of functions~$F_k : M \rightarrow \F$
which is even a minimizing sequence because~$\Sact[F_k] \rightarrow 0$.
In all these examples we satisfy the identity constraint~(C2).

\begin{Example}(Divergent series with bounded action) \label{ex26}
\em We let~$n=1$, $f=2$ and $M=\{1,\ldots 4\}$ with the normalized counting measure.
For any given parameter~$\tau \in \R$ we choose the function~$F : M \rightarrow \F$ by
\[ F(1) = \begin{pmatrix} 4 & 0 \\ 0 & 0 \end{pmatrix} ,\quad
F(2) = \begin{pmatrix} 0 & 0 \\ 0 & 4 \end{pmatrix} \,\quad
F(3) = -F(4) = \tau \begin{pmatrix} 0 & 1 \\ 1 & 0 \end{pmatrix}\:. \]
Obviously, these matrices all have at most one positive and one negative eigenvalue. Furthermore,
the identity constraint~(C2) is satisfied. Using that the matrices~$F(1)$ and~$F(2)$ are of rank one,
we find
\[ \L[A_{11}]  = \frac{1}{2}\: |A_{11}|^2 = 128 =  \L[A_{22}] \:,\qquad
\L[A_{12}] = 0\:. \]
Since the square of the matrix~$F(3)$ is the identity, we obtain~$\L[A_{33}]=0$,
and similarly~$\L[A_{44}]=\L[A_{34}]=0$. The matrix~$F(1) F(3)$ is nilpotent and
and thus~$\L[A_{13}]=0$. Similarly, we find that~$\L[A_{23}]=\L[A_{24}]=\L[A_{34}]=0$.
We conclude that
\[ \Sact[F] = \frac{1}{16} \left( \L[A_{11}]+\L[A_{22}] \right) = 16\:. \]
Hence the action is bounded uniformly in the parameter~$\tau$, but the matrices~$F(3)$
and~$F(4)$ diverge as~$\tau \rightarrow \infty$.
\QEDrem
\end{Example}
This example suggests that the compactness problem could be removed simply by setting the
divergent matrices equal to zero. The next example shows that this simple procedure
does not work, because then the limiting configuration would in general violate the constraints.
\begin{Example} (Violation of the identity constraint) \label{ex27}
\em We let~$n=1$, $f=2$ and $M=\{1,\ldots 3\}$ with the normalized counting measure.
For a given parameter~$\tau>1$ we choose the function~$F : M \rightarrow \F$ as the following linear
combinations of Pauli matrices:
\[ F(1) =  3 \:\frac{\sqrt{1+\tau^2}}{\tau} \:\sigma^3 + 3 \1\:, \qquad
F(2),F(3) = -\frac{3}{2}\:\frac{\sqrt{1+\tau^2}}{\tau} \:\sigma^3 \pm \frac{3}{2}\:
\sqrt{1+\tau^2}\: \sigma^2 \:. \]
A short calculation shows that these matrices all have at most one positive and one negative eigenvalue. Furthermore, the identity constraint~(C2) is satisfied. To compute the action, we first note
that
\[ F(1)^2 = 9 \:\frac{1+2 \tau^2}{\tau^2} \1 + 2\!\cdot\!9\:
\frac{\sqrt{1+\tau^2}}{\tau}\: \sigma^3 
\quad \text{and thus} \quad \L[A_{11}] = 2^3\!\cdot\! 9^2\: \frac{1+\tau^2}{\tau^2}\:. \]
Since the squares of the matrices~$F(2)$ and~$F(3)$ are multiples of the identity,
we find that~$\L[A_{22}]=\L[A_{33}]=0$. Moreover, using the identity for Pauli matrices
\beq \label{Pauliid}
\sigma^i \sigma^j = \delta^{ij} + i \epsilon^{ijk}\: \sigma^k \:,
\eeq
we obtain
\[ A_{12} = c\: \1
-\frac{9}{2}\:\frac{\sqrt{1+\tau^2}}{\tau} \:\sigma^3
+\frac{9}{2}\: \sqrt{1+\tau^2}\: \sigma^2 -
\frac{9i}{2} \: \frac{1+\tau^2}{\tau}\: \sigma^1 \]
with~$c \in \R$. A short calculation using the anti-commutation relations for the Pauli
matrices yields~$(A_{12}-c)^2=0$, and thus the matrix~$A_{12}$ has the eigenvalue~$c$,
with algebraic multiplicity two.
Using the notion of Definition~\ref{causal}, the points~$1$ and~$2$ have spacelike separation,
so that~$\L[A_{12}]=0$. Similarly one verifies that~$\L[A_{13}]=\L[A_{23}]=0$.
We conclude that
\[ \Sact[F] = \frac{1}{9}\: \L[A_{11}] = 72\: \frac{1+\tau^2}{\tau^2}\:. \]

Hence in this example the action is bounded uniformly in~$\tau$, although the matrices~$F(2)$
and~$F(3)$ diverge as~$\tau \rightarrow \infty$. Setting these two matrices to zero, the remaining
matrix~$F(1)$ does converge,
\[ \lim_{\tau \rightarrow \infty} F(1) = 3\, (\sigma^3+\1) \:, \]
but the limiting system no longer satisfies the identity constraint~(C2).
\QEDrem
\end{Example}

\begin{Example} (The two-dimensional Dirac sphere) \label{ex28}
\em We let~$M=S^2 \subset \R^3$ with~$d\mu$ the surface area
measure, normalized such that~$\mu(S^2)=1$. Furthermore, we choose~$n=1$ and~$f=2$.
For a given parameter~$\tau>1$ we introduce the mapping~$F : M \rightarrow \F$ by
\beq \label{FS2}
F(\vec{x}) = \tau\: \vec{x} \vec{\sigma} + \1\:.
\eeq
Then~$\sigma(F(x))=\{1+\tau, 1-\tau\}$, and thus~$F(x)$ has one positive and one negative
eigenvalue. Furthermore, a symmetry argument shows that the identity constraint~(C2) is satisfied.
Using the Pauli identities~\eqref{Pauliid}, one obtains
\[ F(\vec{x}) \,F(\vec{y}) = \left(1+\tau^2\: \vec{x} \vec{y} \right) \1 
+ \tau\: (\vec{x}+\vec{y}) \vec{\sigma} +i \tau^2 \,(\vec{x} \wedge \vec{y}) \vec{\sigma} . \]
A straightforward calculation yields for the eigenvalues of this matrix
\beq \label{leigen}
\lambda_{1\!/\!2} = 1+\tau^2 \cos \vartheta \pm
 \tau \sqrt{1+\cos \vartheta}\: \sqrt{2 - \tau^2 \:(1-\cos \vartheta)} \:,
\eeq
where~$\vartheta$ denotes the angle~$\vartheta$ between~$\vec{x}$ and~$\vec{y}$.
If~$\vartheta$ is sufficiently small, the term~$(1-\cos \vartheta)$ is close to zero, and thus the
arguments of the square roots are all positive. However, if~$\vartheta$ becomes so large that
\[ \vartheta \:\geq\: \vartheta_{\max} := \arccos \!\left( 1-\frac{2}{\tau^2} \right), \]
the argument of the last square root in~\eqref{leigen} becomes negative, so that the~$\lambda_{1\!/\!2}$
form a complex conjugate pair. The calculation
\[ \lambda_1 \lambda_2 = \det(F(\vec{x}) F(\vec{y}))
= \det(F(\vec{x})) \, \det(F(\vec{y})) = (1+\tau)^2 (1-\tau)^2 > 0 \]
shows that if the~$\lambda_{1\!/\!2}$ are both real, then they have the same sign.
Hence the Lagrangian simplifies to
\begin{align*}
\L[A_{xy}] &= \L(\cos \vartheta)
=\frac{(\lambda_1 - \lambda_2)^2}{2}\: \Theta(\vartheta_{\max}-\vartheta) \\
&= 2 \tau^2\: (1+\cos \vartheta) \left( 2 - \tau^2 \:(1-\cos \vartheta) \right) \:
\Theta(\vartheta_{\max}-\vartheta) \:.
\end{align*}
Using this formula in~\eqref{Sdef}, we can carry out the integrals to obtain
\beq \label{SFint}
\Sact[F] = \frac{1}{2} \int_0^{\vartheta_{\max}} \L(\cos \vartheta)\: \sin \vartheta\:
d \vartheta =  4 - \frac{4}{3 \tau^2}\:.
\eeq
Similarly, the functional~$\T$ can be computed to be
\beq \label{TFint}
\T[F] = 4 \tau^2 (\tau^2-2) + 12 -\frac{8}{3 \tau^2} \:.
\eeq
Hence the action~\eqref{SFint} is bounded uniformly in~$\tau$, although the function~$F$, \eqref{FS2},
as well as the functional~$\T$, \eqref{TFint}, diverge as~$\tau \rightarrow \infty$. \QEDrem
\end{Example} \noindent
We remark that this example can be extended to the case of general even~$f$ by
decomposing~$\C^f$ as a direct sum of~$f/2$ copies of~$\C^2$,
choosing~$M = S^2 \times \{1, \ldots, f/2 \}$ and setting
\beq \label{Fex}
F \::\: S^2 \times \{1, \ldots, f/2\} \rightarrow (C^2)^{\frac{f}{2}}\::\:
(\vec{x}, i) \mapsto 0 \oplus \cdots \oplus 0 \oplus 
\underbrace{\left( \tau\: \vec{x} \vec{\sigma} + \1 \right)}_{\text{$i^\text{th}$ summand}}
\oplus 0 \oplus \cdots \oplus 0\:.
\eeq

\begin{Example} (The three-dimensional Dirac sphere) \label{ex29} \em
This example can be regarded as an analog of Example~\ref{ex28},
but in one dimension higher. We introduce the
four $4 \times 4$-matrices
\[ \gamma^\alpha = \begin{pmatrix} \sigma^\alpha & 0 \\ 0 & -\sigma^\alpha \end{pmatrix},
\quad \alpha=1,2,3  \qquad \text{and} \qquad
\gamma^4 = \begin{pmatrix} 0 & \1 \\ \1 & 0 \end{pmatrix} . \]
These are the Dirac matrices of Euclidean~$\R^4$, satisfying the anti-commutation relations
\[ \{\gamma^i, \gamma^j\} = 2 \delta^{ij}\:\1\qquad (i,j=1,\ldots, 4)\:. \]
As is verified either by direct computation or by applying the general theory of Clifford
representations, the Dirac matrices are $\text{\rm{SO}}(4)$-invariant in the sense that for every
rotation~$R \in \text{\rm{SO}}(4)$, there is a unitary matrix~$U \in \text{\rm{SU}}(4)$ such that
\beq \label{SO3}
U \gamma^j U^{-1} = R^j_k \gamma^k\:.
\eeq

We let~$f=4$, $n=2$ and set~$M=S^3$. We introduce the mapping~$F\::\: M \rightarrow \F$ by
\[ F(x) = \sum_{i=1}^4 \tau\: x^i \gamma^i + \1\:. \]
These matrices are $\text{\rm{SO}}(4)$-invariant in the sense that
\[ F(R x) = U F(x) U^{-1} \:, \]
where~$R$ and~$U$ are the transformations in~\eqref{SO3}. Since the unitary transformation~$U$
does not affect the eigenvalues of~$A_{xy}$, we see that~$\L[A_{xy}] = \L[A_{Rx\, Ry}]$.
Thus the Lagrangian will depend only on the angle~$\vartheta$ between the vectors~$x$ and~$y$.
Furthermore, it suffices to compute the Lagrangian for vectors~$x$ and~$y$ for which the
zero component vanishes. But in this case, the eigenvalues of~$A_{xy}$ are calculated
exactly as in Example~\ref{ex28} above. Thus the eigenvalues are again given by~\eqref{leigen},
but now each eigenvalue appears with algebraic multiplicity two. We conclude that
\begin{align*}
\L[A_{xy}] &= \L(\cos \vartheta)
= (\lambda_1 - \lambda_2)^2\: \Theta(\vartheta_{\max}-\vartheta) \\
&= 4 \tau^2\: (1+\cos \tau) \left( 2 - \tau^2 \:(1-\cos \vartheta) \right) \:
\Theta(\vartheta_{\max}-\vartheta) \:.
\end{align*}
Inserting this Lagrangian in~\eqref{Sdef},
\[ \Sact[F] = \frac{2}{\pi} \int_0^{\vartheta_{\max}}
\L(\cos \vartheta)\: \sin^2 \vartheta\:d \vartheta\:, \]
we obtain the same integral as in~\eqref{SFint}, except that the integrand contains
an additional factor~$\sin \vartheta$. Due to this extra factor, the action decays
for large~$\tau$,
\[ \Sact[F] = \frac{512}{15 \pi}\: \frac{1}{\tau} + {\mathscr{O}}(\tau^{-2})\:. \]
Thus setting~$F_k =F|_{\tau=k}$, we have constructed a divergent minimizing sequence.
\QEDrem
\end{Example} \noindent
We finally remark that this example generalizes similar to~\eqref{Fex} to larger~$f$,
provided that~$f$ is divisible by four.

\subsection{The Moment Measures and the Possibility of Bubbling} \label{sec23}
In this section we discuss the main difficulties in proving Theorems~\ref{thm4} and~\ref{thm5},
and we will explain the methods for resolving these difficulties. The obvious starting point is a minimizing
sequence~$F_k \in \M$. In the setting of Theorem~\ref{thm5}, we know that~$\T$ is uniformly bounded,
\beq \label{Tbound}
\T[F_k] \leq C \qquad \text{for all $k \in \N$}.
\eeq
In the setting of Theorem~\ref{thm4}, the estimate~\eqref{Les} shows that~\eqref{Tbound} again
holds if we set
\[ C = \left[1 + \min(\nu, 0) \left( 1-\frac{1}{2n}\right) \right]^{-1} \max_{k \in \N}
(\T[F_k] + \nu \Sact[F_k]) < \infty \:. \]
In view of the freedom to act by isomorphisms of the measure space~$(M, \mu)$ discussed
after~\eqref{Udef}, it is preferable to work again instead of the mapping~$F$
with the corresponding measure~$\rho$ on~$\F$ as defined by~\eqref{rhodef}. Similar to~\eqref{Srho},
we can write the functionals~$\Sact$ and~$\T$ in terms of~$\rho$,
\beq \label{STrho}
\Sact(\rho) = \iint_{\F \times \F} \L[p \,q]\: d\rho(p) \,d\rho(q) \:,\qquad
\T(\rho) = \iint_{\F \times \F} |p \, q|^2\: d\rho(p) \,d\rho(q) \:.
\eeq
Moreover, the identity
\beq \label{cint}
\int_M F(x) \: d\mu(x) = \int_\F p\: d\rho(p)
\eeq
allows us to also express the constraints~(C1) and~(C2) in terms of~$\rho$.

The main complication compared to the setting of Chapter~\ref{sec1} is that~$\F$ is no
longer compact, and thus we need to control the support of the measures~$\rho_k$
in order to ensure that the limiting measure
$\rho=\lim_{k \rightarrow \infty} \rho_k$ again has total volume one.
Moreover, it is no longer obvious that the functionals in~\eqref{STrho} or
the integral~\eqref{cint} converge in the limit~$k \rightarrow \infty$. At this point,
it is helpful to observe that the integrand in~\eqref{cint} is homogeneous in~$p$ of
degree one, whereas the integrands in~\eqref{STrho} are homogeneous of degree two
in both~$p$ and~$q$. This allows us to express these functions in terms of so-called
moment measures, which we now introduce. 
\begin{Def} \label{defmm}
Let~$\K$ be the compact set
\[ \K = \{ p \in \F \text{ with } \|p\|=1 \} \cup \{0\} \:. \]
For a given measure~$\rho$ on~$\F$ we define the measurable sets of~$\K$ by the
requirement that the sets~$\R^+ \Omega = \{ \lambda p \:|\: \lambda \in \R^+, p \in \Omega\}$
and~$\R^- \Omega$ should be $\rho$-measurable in~$\F$. We introduce the measures~$\m^{(0)}$, 
$\m^{(1)}$ and~$\m^{(2)}$ by
\begin{align}
\m^{(0)}(\Omega) &= \frac{1}{2}\: \rho \big(\R_+ \Omega \setminus \{0\} \big) 
+ \frac{1}{2}\: \rho \big( \R_- \Omega \setminus \{0\} \big)
+ \rho \big( \Omega \cap \{0\} \big) \label{m0def} \\
\m^{(1)}(\Omega) &= \frac{1}{2} \int_{\R^+ \Omega} \|p\| \,d\rho(p) \:-\:
\frac{1}{2} \int_{\R^- \Omega} \|p\| \,d\rho(p) \label{m1def} \\
\m^{(2)}(\Omega) &= \frac{1}{2} \int_{\R_+ \Omega} \|p\|^2 \,d\rho(p) \:+\:
\frac{1}{2} \int_{\R_- \Omega} \|p\|^2 \,d\rho(p)
\:. \label{m2def}
\end{align}
The measure~$\m^{(l)}$ is referred to as the $l^\text{th}$ {\bf{moment measure}}.
\end{Def} \noindent
In terms of the moment measures, the normalization~$\rho(\F)=1$ becomes
\beq \label{m0}
\m^{(0)}(\K) = 1 \:,
\eeq
whereas the relations~\eqref{STrho} and~\eqref{cint} can be written as
\begin{align}
\Sact(\rho) &= \iint_{\K \times \K} \L[p \,q]\: d\m^{(2)}(p) \,d\m^{(2)}(q) \label{Sm2} \\
\T(\rho) &= \iint_{\K \times \K} |p \, q|^2\:  d\m^{(2)}(p) \,d\m^{(2)}(q) \label{Tm2} \\
\int_\K p\: d\m^{(1)} &= \int_\F p\, d\rho = \int_M F(x)\, d\mu(x) \:. \label{m1}
\end{align}

Working with the~$\m^{(l)}$ has the advantage that they are measures on the
{\em{compact}} space~$\K$. We also learn that two measures~$\rho$ and~$\tilde{\rho}$
whose moment measures coincide yield the same values for the functionals~$\Sact$ and~$\T$ as well
as for the integral~\eqref{m1} entering the constraints. Therefore, it is useful
to consider two measures as being {\em{equivalent}} if their
moment measures~$\m^{(0)}$, $\m^{(1)}$ and~$\m^{(2)}$ coincide.
Using this notion, we have a large freedom to modify the measures~$\rho_k$ within
the equivalence class defined by its moment measures~$(\m^{(0)}_k, \m^{(1)}_k, \m^{(2)}_k)$.
This freedom is indeed a problem for proving convergence of the measures~$\rho_k$, as the following
one-dimensional analog shows.
\begin{Example} (Discontinuous moments) \label{ex211}
\em We consider on~$\R$ the family of measures
\beq \label{rhoex}
\rho_\tau = \frac{3}{\tau}\: \delta_0 +
\frac{\tau-4}{\tau-1}\: \delta_1 + \frac{3}{\tau^2-\tau} \delta_\tau \:, \qquad
\tau >1\:,
\eeq
where~$\delta_x$ denotes the Dirac measure supported at~$x$. These measures all have the
same moments
\beq \label{momex}
\rho_\tau(\R) = 1\:, \qquad \int_\R x\: d\rho_\tau(x) = 1 \:,\qquad \int_\R x^2\: d\rho_\tau(x) = 4\:,
\eeq
and are thus all equivalent in the above sense.

In the limit~$\tau \rightarrow \infty$, the measures~$\rho_\tau$ converge in the weak
$C^0(\R)^*$-topology\footnote{For clarity we point out that by~$C^0$ we always mean the
closure of the compactly supported continuos functions with respect to the $\sup$-norm.
Thus $C^0(\R)$ is the space of continuous functions~$f$ with~$\lim_{|x| \rightarrow \infty} |f|=0$.}
to the measure~$\delta_1(x)\, dx$, having the moments
\[ \delta_1(\R) = 1\:, \qquad \int_\R x\: \delta_1(x)\, dx = 1 \:,\qquad \int_\R x^2\: \delta_1(x)\, dx = 1\:. \]
Hence the total volume is preserved in the limit, and also the first moment is continuous.
But the second moment jumps discontinuously as~$\tau \rightarrow \infty$.
\QEDrem
\end{Example}
This example reveals the undesirable fact that the limit~$\rho = \lim_k \rho_k$ in the $C^0(\F)^*$-topology may depend on how the representatives~$\rho_k$ of the
corresponding moment measures~$(\m^{(0)}_k, \m^{(1)}_k, \m^{(2)}_k)$ are chosen.
In order to bypass this problem, we shall work exclusively with the moment measures.
At the very end, we shall then construct a suitable representative~$\rho$ of the limiting moment
measures. A key step for making this method work is the following a-priori estimate.

\begin{Lemma} \label{lemma212} There is a constant~$\varepsilon=\varepsilon(f,n)>0$ such that for
every measure~$\rho$ on~$\F$ the corresponding moment measures (see Definition~\ref{defmm})
satisfy for all measurable~$\Omega \subset \K$ the following inequalities:
\begin{align}
\left| \m^{(1)}(\Omega) \right|^2 &\leq \m^{(0)}(\Omega)\:\m^{(2)}(\Omega) \label{m1es} \\
\m^{(2)}(\K) &\leq\; \frac{\sqrt{\T(\rho)}}{\varepsilon}\:. \label{m2es}
\end{align}
\end{Lemma}
\Proof The inequality~\eqref{m1es} follows immediately from H\"older's inequality,
\[ \big| 2 \m^{(1)}(\Omega) \big|^2 \leq  \left( \int_{\R \Omega} \|p\| \,d\rho(p) \right)^2
\leq \rho(\R \Omega) \int_{\R \Omega} \|p\|^2 \,d\rho(p)
\leq 4 \m^{(0)}(\Omega)\: \m^{(2)}(\Omega) \:. \]

To prove~\eqref{m2es}, we introduce the mapping
\[ \phi \::\: \K \times \K \rightarrow \R \::\: (p,q) \mapsto |p \, q|\:. \]
Clearly, $\phi$ is continuous and
\[ \phi(p,p) = |p^2| = \Tr(p^2) = \|p\|^2 = 1\:. \]
Thus every point~$r \in \K$ has a neighborhood $U(r) \subset \K$ with
\beq \label{Uineq}
\phi(p,q) \geq \frac{1}{2} \qquad \text{for all $p,q \in U(r)$}\:.
\eeq
Since~$\K$ is compact, there is a finite number of points~$r_1,\ldots, r_N$
such that the corresponding sets~$U_i:=U(r_i)$ cover~$\K$. Due to the additivity property
of measures, there is an index~$i \in \{1, \ldots, N\}$ such that
\beq \label{mui}
\m^{(2)}(U_i) \geq \frac{\m^{(2)}(\K)}{N}\:.
\eeq

We write~$\T$ in the form~\eqref{Tm2} and apply~\eqref{Uineq} as well
as~\eqref{mui} to obtain
\[ \T(\rho) \geq \iint_{U_i \times U_i} |p \, q|^2\: d\m^{(2)}(p) \,d\m^{(2)}(q)
\geq \frac{1}{2}\: \m^{(2)}(U_i)^2 \geq
\frac{\m^{(2)}(\K)^2}{2 N^2}\:. \]
Setting~$\varepsilon=1/(\sqrt{2} N)$, the result follows.
\QED

In view of this lemma and the a-priori bound~\eqref{Tbound},
we know that the moment measures are uniformly bounded measures on a
compact space~$\K$.
Thus, exactly as in Section~\ref{sec12}, we can apply the Banach-Alaoglu theorem and the Riesz
representation theorem to conclude that for a suitable subsequence of the~$F_k$ (which for simplicity we
denote again by~$F_k$), these measures converge in the~$C^0(\K)^*$-topology to
regular Borel measures,
\[ \m_k^{(l)} \rightarrow \m^{(l)} \qquad (l \in \{0,1,2\}) , \]
which again have the properties~\eqref{m0}, \eqref{m1es} and~\eqref{m2es}.

We next consider the Radon-Nikodym decompositions of~$\m^{(1)}$ and~$\m^{(2)}$ with
respect to~$\m^{(0)}$ (cf.~\cite[Section~31]{halmosmt}),
\[ d\m^{(l)} = f^{(l)}\, d\m^{(0)} + d\m^{(l)}_\text{sing} \qquad \text{with~$f^{(l)}
\in L^1(\K, d\m^{(0)})$}\qquad (l=1,2)\:, \]
where the measures~$d\m^{(l)}_\text{sing}$ are singular with respect to~$d\m^{(0)}$.
Evaluating~\eqref{m1es} for any~$\Omega$ in the support of~$d\m^{(1)}_\text{sing}$,
the right side vanishes, and thus~$\m^{(1)}_\text{sing}=0$. Furthermore, the
inequality~\eqref{m1es} implies that~$|f^{(1)}|^2 \leq f^{(2)}$. In particular, we
conclude that~$f^{(1)}$ even lies in~$L^2(\K, d\m^{(0)})$. Setting~$f=f^{(1)}$
and~$d\n^{(2)} = (f^{(2)}-|f|^2)\, d\m^{(0)} + d\m^{(2)}_\text{sing}$,
we obtain the decomposition
\beq \label{RN}
d\m^{(1)} = f\, d\m^{(0)} \:,\qquad
d\m^{(2)} = |f|^2\, d\m^{(0)} + d\n\:,
\eeq
where~$f \in L^2(\K, d\m^{(0)})$, and~$\n$ is a positive measure which need not be
absolutely continuous with respect to~$\m^{(0)}$. From the definition~\eqref{m1def}
it is clear that~$f$ is odd in the sense that
\beq \label{fodd}
f(-p) = -f(p) \quad \text{for all~$p \in \K$}\:.
\eeq

The remaining task is to represent the limiting moment measures~$\m^{(l)}$ in~\eqref{RN} by
a function~$F \in \M$ and a corresponding measure~$\rho$ on~$\F$.
Unfortunately, there is the basic problem that such a measure~$\rho$ can exist only
if~$\m^{(2)}$ is absolutely continuous with respect to~$\m^{(0)}$, as the following consideration shows.
Assume conversely that~$\m^{(2)}$ is not absolutely continuous with respect to~$\m^{(0)}$.
Then there is a measurable set~$\Omega \subset \K$ with~$\m^{(0)}(\K)=0$
and~$\m^{(2)}(\K) \neq 0$.
Assume furthermore that there is a measure~$\rho$ on~$\F$ which represents the limiting
moment measures in the sense that~\eqref{m0def}--\eqref{m2def} hold.
From~\eqref{m0def} we conclude that the set~$\R \Omega \subset \F$
has $\rho$-measure zero. But then the integral~\eqref{m2def} also vanishes, a contradiction.

This problem can also be understood in terms of the limiting sequence~$\rho_k$.
We cannot exclude that there is a star-shaped region~$\R \Omega \subset \F$
such that the measures~$\rho_k(\R \Omega)$ tend to zero, but the corresponding moment
integrals~\eqref{m2def} have a non-zero limit. Using a notion from the calculus of variations
for curvature functionals, we refer to this phenomenon as the possibility of {\em{bubbling}}.
This bubbling effect is illustrated by the following example.

\begin{Example} (Bubbling) \label{ex213} \em
We choose~$f=2$, $n=1$ and again use the Pauli representation~\eqref{Paulirep2}.
Furthermore, we let~$M=[0, 1)$ with~$\mu$ the Lebesgue measure.
For any parameters~$\kappa \geq 0$ and~$\varepsilon \in (0, \frac{1}{2})$,
we introduce the function $F_\varepsilon : M \rightarrow \F$ by
\[ F_\varepsilon(x) = \displaystyle \frac{1}{1-2 \varepsilon} \times \left\{ \begin{array}{ll}
-\kappa \,\varepsilon^{-\frac{1}{2}}\: \sigma^3 & \text{if $x \leq \varepsilon$} \\[.1em]
 \1+ \sigma^1 \,\cos(\nu x) + \sigma^2 \,\sin(\nu x) \quad& \text{if $\varepsilon < x
\leq 1-\varepsilon $} \\[.1em]
\kappa \,\varepsilon^{-\frac{1}{2}}\: \sigma^3 & \text{if $x > 1-\varepsilon\:,$}
\end{array} \right. \]
where we set~$\nu = 2 \pi/(1-2 \varepsilon)$.
The corresponding measure~$\rho_\varepsilon$ on~$\F$ has the following properties. On the set
\[ S := \{ \1 + v^1 \sigma^1 + v^2 \sigma^2 \text{ with } (v^1)^2 + (v^2)^2 = 1 \} \:, \]
which can be identified with a circle~$S^1$, $\rho_\varepsilon$ is
a multiple of the Lebesgue measure. Moreover, $\rho_\varepsilon$ is supported at the
two points
\beq \label{pmdef}
p_\pm := \pm \frac{\kappa\, \varepsilon^{-\frac{1}{2}}}{1-2 \varepsilon}\: \sigma^3 \qquad
\text{with} \qquad \rho(\{p_+\}) = \rho(\{p_-\}) = \varepsilon \:.
\eeq

A short calculation shows that the identity constraint~(C2) is satisfied.
Furthermore, the separations of the points~$p_+$ and~$p_-$ from each other and from~$S$
are either spacelike or just in the boundary case between spacelike and timelike. 
Thus for computing the action, we only need to take into account pairs~$(x,y)$ of points on~$S$.
A straightforward computation yields
\beq \label{STform}
\Sact(\rho_\varepsilon) = \frac{3}{(1-2 \varepsilon)^2} \:,\qquad
\T(\rho_\varepsilon) = \frac{6}{(1-2 \varepsilon)^2}  
+ \frac{16 \kappa^2}{(1-2 \varepsilon)^3} + \frac{16 \kappa^4}{(1-2 \varepsilon)^4}\:.
\eeq

Let us consider the limit~$\varepsilon \searrow 0$. From~\eqref{STform} we see that
the functionals~$\Sact$ and~$\T$ converge,
\beq \label{STval}
\lim_{\varepsilon \searrow 0} \Sact = 3\:, \qquad
\lim_{\varepsilon \searrow 0} \T = 6 + 16\, (\kappa^2 + \kappa^4)\:.
\eeq
Moreover, there are clearly no convergence problems on the set~$S$.
Thus it remains to consider the situation at the two points~$p_\pm$, \eqref{pmdef},
which move to infinity as~$\varepsilon$ tends to zero. These two points enter the moment
measures only at the corresponding normalized points~$\hat{p}_\pm = p_\pm/\|p_\pm\| \in \K$.
A short calculation shows that the limiting moment measures~$\m^{(l)}=\lim_{\varepsilon \searrow 0}
\m^{(l)}_\varepsilon$ satisfy the relations
\[ \m^{(0)}(\{\hat{p}_\pm\}) = 
\m^{(1)}(\{\hat{p}_\pm\}) = 0 \qquad \text{but} \qquad
\m^{(2)}(\{\hat{p}_\pm\}) = \kappa^2 > 0 \:. \]
Hence~$\m^{(2)}$ is indeed {\em{not}} absolutely continuous with respect to~$\m^{(0)}$.

In order to clarify the connection to earlier examples, we point out that, in contrast
to Examples~\ref{ex26} and~\ref{ex27}, the ``bubbles'' have $\rho$-measure zero,
so that the total measure is preserved in the limit. Nevertheless, the bubbles carry
non-zero second moments.
In Example~\ref{ex211} the situation is simpler because the moments~\eqref{momex}
are independent of~$\tau$, whereas in the present example the singularity of the
moment measure appears only in the limit~$\varepsilon \searrow 0$.
In particular, the moment measures~\eqref{momex} can be represented by
a measure~$\rho$ (for example by choosing~$\rho$ according
to~\eqref{rhoex} with~$\tau=2$),
whereas in the present example the corresponding representation~\eqref{m0def}--\eqref{m2def}
cannot be given.

To avoid misunderstandings, we also point out that this example does {\em{not}} show that
bubbling really occurs for minimizing sequences, because we do not know whether the
family~$(F_\varepsilon)_{0<\varepsilon<1/2}$ is minimizing. But at least,
our example shows that bubbling makes it possible to arrange arbitrary large values of~$\T$,
without increasing the action~$\Sact$ (see~\eqref{STval} for large~$\kappa$).
In particular, for large prescribed~$\T$, the action here is strictly smaller
than in Example~\ref{ex28} (cf.~\eqref{SFint} and~\eqref{TFint}).
\QEDrem
\end{Example}

In order to handle possible bubbling phenomena, it is important to observe that the second
moment measure does not enter the constraints~(C1) or~(C2) (see~\eqref{cint} and~\eqref{m1}).
Therefore, by taking out the term~$d\n$ in~\eqref{RN} we {\em{decrease}} the
functionals~$\Sact$ and~$\T$ (see~\eqref{Sm2} and~\eqref{Tm2}), without affecting the
constraints~(C1) or~(C2). The following lemma allows us to remove the term~$d\n$ in~\eqref{RN}.

\begin{figure}
\begin{center}
\begin{picture}(0,0)%
\includegraphics{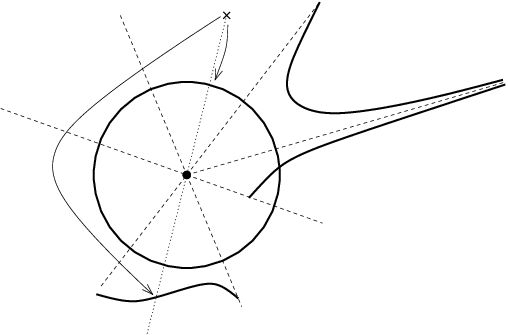}%
\end{picture}%
\setlength{\unitlength}{1865sp}%
\begingroup\makeatletter\ifx\SetFigFontNFSS\undefined%
\gdef\SetFigFontNFSS#1#2#3#4#5{%
  \reset@font\fontsize{#1}{#2pt}%
  \fontfamily{#3}\fontseries{#4}\fontshape{#5}%
  \selectfont}%
\fi\endgroup%
\begin{picture}(8585,5670)(3589,-6373)
\put(7876,-5146){\makebox(0,0)[lb]{\smash{{\SetFigFontNFSS{11}{13.2}{\rmdefault}{\mddefault}{\updefault}${\mathcal{K}}$}}}}
\put(7446,-1666){\makebox(0,0)[lb]{\smash{{\SetFigFontNFSS{11}{13.2}{\rmdefault}{\mddefault}{\updefault}$\pi_{\mathcal{K}}$}}}}
\put(5406,-3016){\makebox(0,0)[lb]{\smash{{\SetFigFontNFSS{11}{13.2}{\rmdefault}{\mddefault}{\updefault}$f=0$}}}}
\put(7166,-4436){\makebox(0,0)[lb]{\smash{{\SetFigFontNFSS{11}{13.2}{\rmdefault}{\mddefault}{\updefault}$f=0$}}}}
\put(5436,-3956){\makebox(0,0)[lb]{\smash{{\SetFigFontNFSS{11}{13.2}{\rmdefault}{\mddefault}{\updefault}$f<0$}}}}
\put(7586,-1066){\makebox(0,0)[lb]{\smash{{\SetFigFontNFSS{11}{13.2}{\rmdefault}{\mddefault}{\updefault}$p$}}}}
\put(3916,-3616){\makebox(0,0)[lb]{\smash{{\SetFigFontNFSS{11}{13.2}{\rmdefault}{\mddefault}{\updefault}$\pi^f$}}}}
\put(10186,-3286){\makebox(0,0)[lb]{\smash{{\SetFigFontNFSS{11}{13.2}{\rmdefault}{\mddefault}{\updefault}${\mathcal{K}}^f = \{p \:|\: \|p\| = f(\pi_{\mathcal{K}}(p))\}$}}}}
\end{picture}%
\caption{Example for the sets~$\K$, $\K^f$ and the projections~$\pi_\K$, $\pi^f$.}
\label{fig3}
\end{center}
\end{figure}
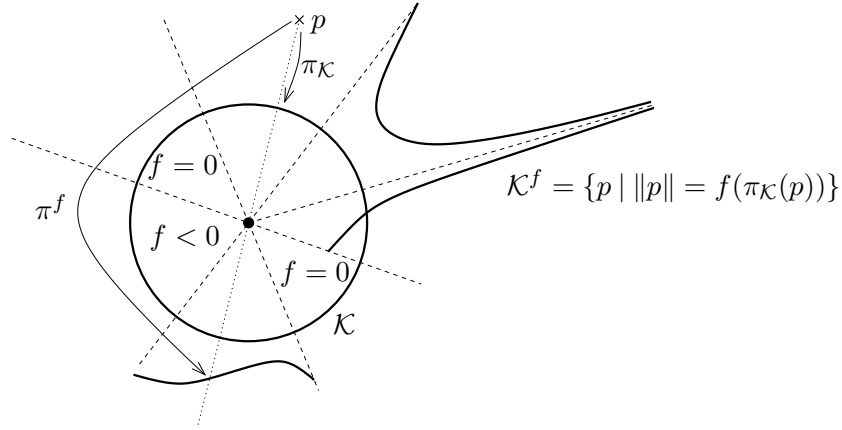

\begin{Lemma} \label{lemma214}
For every function~$F \in \M$ having the moment measures~\eqref{RN}
there is a function~$\tilde{F} \in \M$ whose moment measures~$\tilde{\m}^{(l)}$ are given by
\beq \label{tmmom}
\tilde{\m}^{(0)} = \m^{(0)}\:, \qquad
d\tilde{\m}^{(1)} =  f\, d\m^{(0)} \:,\qquad
d\tilde{\m}^{(2)} = |f|^2\, d\m^{(0)}\:.
\eeq
\end{Lemma}
\Proof
We first introduce the projection
\[ \pi_\K \::\: \F \rightarrow \K \::\: p \mapsto \left\{
\begin{array}{cl} 0 & \text{if $p=0$} \\[0.3em]
\displaystyle \frac{p}{\|p\|} & \text{if $p \neq 0$}\:. \end{array} \right. \]
For the function~$f$ in~\eqref{RN} we introduce the set~$\K^f \subset \F$ and the corresponding
projection~$\pi^f$ by
\begin{align*}
\K^f &= \left\{ f(p) \,p \text{ with } p \in \K \right\} \subset \F \\
\pi^f &\::\: \F \rightarrow \K^f \::\: p \mapsto
f(\pi_\K(p))\: \pi_\K(p)\:.
\end{align*}
These definitions are illustrated in Figure~\ref{fig3}.
Finally, we introduce the function~$\tilde{F}$ by
\beq \label{tFdef}
\tilde{F} \::\: M \rightarrow \K^f \subset \F \::\: x \mapsto \pi^f(F(x)) \:.
\eeq

We decompose any measurable set~$\Omega \subset \K$ as
$\Omega = \Omega_0 \dot{\cup} \Omega_+ \dot{\cup} \Omega_-$, where
\[ \Omega_0 = \{p \in \Omega \:|\: f(p)=0\} \:,\quad
\Omega_+ = \{p \in \Omega \:|\: f(p) > 0\} \:,\quad
\Omega_- = \{p \in \Omega \:|\: f(p) < 0\} \:. \]
Since~$f$ is odd~\eqref{fodd}, we know that~$f(0)=0$ and thus~$0 \in \Omega_0$.
Using the definition of~$\tilde{F}$, \eqref{tFdef}, it follows that
\[ \tilde{\rho}(\Omega_0) = \rho(\{0\}) \:,\qquad \tilde{\rho}(\R^+ \Omega_+) = \rho(\Omega_+)
\:,\qquad \tilde{\rho}(\R^+ \Omega_-) = 0\:. \]
Moreover, using~\eqref{m0def} together with the fact
that the measure~$\tilde{\rho}$ is supported on~$\K^f$, we find
\[ \int_{\R_+ \Omega_+} \|p\|^l\: d\tilde{\rho}(p) = 2 \int_{\Omega_+} |f(p)|^l\: d\m^{(0)}(p)
\qquad (l \in \{0,1,2\})\:. \]
Similarly,
\[ \int_{\R_- \Omega_-} \|p\|^l\: d\tilde{\rho}(p) = 2 \int_{\Omega_-} |f(p)|^l\: d\m^{(0)}(p) \:. \]
Using these relations, we can compute the moment measures corresponding to~$\tilde{\rho}$ by
\begin{align*}
\tilde{\m}^{(0)}(\Omega) &= \frac{1}{2}\: \tilde{\rho} (\R \Omega_+)
+ \frac{1}{2}\: \tilde{\rho} (\R \Omega_-) + \tilde{\rho}(\Omega_0) \\
&= \m^{(0)}(\Omega_+) + \m^{(0)}(\Omega_-) + \m^{(0)}(\Omega_0) = \m^{(0)}(\Omega) \\
\tilde{\m}^{(1)}(\Omega) &= \frac{1}{2} \int_{\R_+ \Omega_+} \|p\| \,d\tilde{\rho}(p) -
\frac{1}{2} \int_{\R_- \Omega_-} \|p\| \,d\tilde{\rho}(p) \\
&= \int_{\Omega_+} f(p) \,d\m^{(0)}(p) +\int_{\Omega_-} f(p) \,d\m^{(0)}(p)
= \int_{\Omega_+ \cup \Omega_-} f(p)\: d\m^{(0)}(p) \\
\tilde{\m}^{(2)}(\Omega) &= \frac{1}{2} \int_{\Omega_+ \cup \Omega_-} \|p\|^2 \,d\tilde{\rho}(p)
= \int_{\Omega_+ \cup \Omega_-} |f(p)|^2\: d\m^{(0)}(p) \:.
\end{align*}
This completes the proof.
\QED
Applying this lemma to our minimizing sequence, we shall obtain a new minimizing
sequence~$\tilde{F}_k$. The next lemma shows that the limiting moment measures of
this new sequence can indeed be represented by a function~$F \in \M$.
\begin{Lemma} \label{lemma215} Let~$F_k \in \M$ be a sequence of functions such that the
corresponding moment measures~$\m^{(l)}_k$, $l \in \{0,1,2\}$, converge in the
$\C^0(\K)^*$-topology to moment measures~$\m^{(l)}$ having the Radon-Nikodym representation
\beq \label{RNdec}
d\m^{(1)} = f \,d\m^{(0)}\:,\qquad d\m^{(2)} = |f|^2 \,d\m^{(0)}
\eeq
with~$f \in L^2(\K, d\m^{(0)})$.
Then there is a function~$F \in L^2(M, \F, d\mu)$ such that the corresponding moment measures
(as defined by~\eqref{m0def}--\eqref{m2def} and~\eqref{rhodef}) coincide with the
moment measures in~\eqref{RNdec}.
\end{Lemma}
\Proof If the moment measures corresponding to a function~$F \in \M$ coincide with~\eqref{RNdec}, it
follows from Definition~\ref{defmm} that
\[ \int_M \|F\|^2 \,d\mu = \int_{\F} \|p\|^2 \,d\rho(p) = \m^{(2)}(\K) 
= \|f\|^2_{L^2(\K, d\m^{(0)})} < \infty\:, \]
proving that the function is square integrable, $F \in L^2(M, \F, d\mu)$.

Using the same notation as in the proof of Lemma~\ref{lemma214}, we define
the measure~$\rho$ on~$\F$ by
\[ \rho(\{0\}) = \m^{(0)}(\{0\}) \qquad \text{and} \qquad
\rho(U) = 2 \m^{(0)}\! \left( \pi_\K (U \cap \K^f ) \right) \text{ if $0 \not \in U$}. \]
Similar as in the proof of Lemma~\ref{lemma214} one verifies that the moment measures
corresponding to~$\rho$ indeed coincide with the moment measures in~\eqref{RNdec}.
Thus it remains to represent~$\rho$ by a function~$F \in \M$.

As in the proof of Lemma~\ref{lemma14}, we decompose the measure space~$(M, \mu)$
into an atomic part~$(M^\text{d}, \mu^\text{d})$ and a non-atomic part~$(M^\text{c}, \mu^\text{c})$
(see~\eqref{Mdec}). By modifying the mappings~$F_k$ on sets of measure zero we can again
arrange that for every atom~$A \in \mathfrak{A}$, the set~$F_k(A)$ consist of only one point.
The estimate
\beq \label{nobubble}
\|F_k(A)\|^2\, \mu^\text{c}(A) \leq \int_M \|F_k(x)\|^2 \, d\mu = \m^{(2)}_k(\K)
\xrightarrow{k \rightarrow \infty} \m^{(2)}(\K)
\eeq
shows that for any~$A \in \mathfrak{A}$, the sequence~$F_k(A)$ is bounded in~$\F$.
Thus, just as in the proof of Lemma~\ref{lemma14}, there is a subsequence of the~$F_k$
such that the sequence~$F_k(A)$ converges for every~$A \in \mathfrak{A}$.
Again subtracting the resulting limit measure according to~\eqref{rhocdef}, we obtain
a positive~$\rho^\text{c}$. Thus it remains to construct a function~$g : M^\text{c} \rightarrow \F$
which satisfies~\eqref{gcond}.

The method of Lemma~\ref{lemma14} does not immediately apply because the
support of~$\rho$ does not need to be bounded. But we can use the following standard
exhaustion argument. We introduce the sets~$\F_L$, $L \in \N_0$ by
\[ \F_L = \{p \in \F \text { with } L \leq \|p\| < L+1\}\:. \]
Clearly, the $\F_L$ form a partition of~$\F$. Using~\eqref{decomp}, we can iteratively
construct a partition~$M_L$ of~$M^\text{c}$ such that~$\mu^\text{c}(M_L) = \rho(\F_L)$.
Exactly as in the proof of Lemma~\ref{lemma14} we can construct a functions~$g_L \::\: M_L
\rightarrow \F_L$ with the property
\[ \rho^\text{c}(\Omega) = \mu^\text{c}(g_L^{-1}(\Omega))  \quad
\text{for every Borel set~$\Omega \subset \F_L$}\:. \]
Then the function~$g$ defined by~$g|_{M_L}=g_L$ has the required property~\eqref{gcond}.
\QED

\subsection{Existence Proofs} \label{sec24}
We can now complete the proofs of the existence theorems stated in Section~\ref{sec21}.
Theorem~\ref{thm3} can be proved with the same methods as Theorem~\ref{thm1}.
\begin{proof}[Proof of Theorem~\ref{thm3}.]
The set~$\F_\text{(C3)}$ of all matrices satisfying the constraint~(C3) is a compact
manifold. For a given minimizing sequence~$(F_k)_{k \in \N}$ the corresponding
measures~$\rho_k$ defined by~\eqref{rhodef} are supported on~$\F_\text{(C3)}$.
Exactly as in the proof of Lemma~\ref{lemmameasure}, a subsequence of the $\rho_k$
converges in the~$C^0(\F_\text{(C3)})^*$-topology to a positive normalized Borel measure
on~$\F_\text{(C3)}$. Since the integrands in~\eqref{STrho} and~\eqref{cint} are continuous
in~$p$ and~$q$, the action converges and the constraints~(C1) or~(C2) (if considered)
are preserved in the limit.
As in the proof of Lemma~\ref{lemma14} we finally represent~$\rho$ by a function~$F \in \M$.
\QED
For the remaining proofs we need to use the methods of Section~\ref{sec23}.
\begin{proof}[Proof of Theorems~\ref{thm4} and~~\ref{thm5}.]
Let~$(F_k)_{k \in \N}$ be a minimizing sequence. According to Lemma~\ref{lemma214},
to any~$F_k$ we can associate a function~$\tilde{F}_k \in \M$. Using~\eqref{tmmom}
and~\eqref{RN} in~\eqref{Sm2}--\eqref{m1}, we see that
\[ \Sact[\tilde{F}_k] \leq \Sact[F_k]\:, \quad \T[\tilde{F}_k] \leq \T[F_k] \qquad \text{and} \qquad
\int_{M} \tilde{F}_k(x)\: d\mu(x) = \int_{M} F_k(x)\: d\mu(x)\:. \]
Hence replacing the~$F_k$ by the~$\tilde{F}_k$, we obtain a new minimizing sequence
which still satisfies all the constraints. As in the proof of Lemma~\ref{lemmameasure},
the Banach-Alaoglu theorem and the Riesz representation theorem yield that for a subsequence
of the~$F_k$ (denoted again by~$F_k$), the corresponding moment measures converge
in the $C^0(\K)^*$-topology to bounded regular Borel measures~$\m^{(l)}$, $l \in \{0,1,2\}$.
The property~\eqref{tmmom} of the~$F_k$ yields that the measures~$\m^{(1)}$
and~$\m^{(2)}$ have the Radon-Nikodym representation~\eqref{RNdec}
with~$f \in L^2(\K, d\m^{(0)})$. According to Lemma~\ref{lemma215}, we can represent
the moment measures by a function~$F \in L^2(M, \F, d\mu)$, being the desired minimizer.
\QED

\subsection{Remarks and Open Problems}
We conclude this chapter by a few remarks and a brief discussion of
open problems. We first note that Theorems~\ref{thm4} and~\ref{thm5} remain valid
if other constraints are imposed. For example, one could prescribe only
some of the $2n$ non-trivial eigenvalues of~$F(x)$, or one could prescribe the trace of~$F(x)$.
Such generalizations are straightforward and shall not be considered here.
More interesting are the following open problems:
\begin{itemize}
\item[(A)] It is not clear whether Theorem~\ref{thm5} remains valid if the
constraint $\T[F] \leq C$ is replaced by~$\T[F]=C$. In other words, does the bubbling phenomenon discussed in Section~\ref{sec23} really occur for minimal sequences,
or does it only reflect a shortcoming of our method of proof?
We again point out that Example~\ref{ex213} does not prove bubbling, because it is
unknown whether the constructed family~$F_\varepsilon$ is minimal.
\item[(B)] It is an open problem whether Theorem~\ref{thm5} holds without the constraint~$\T[F] \leq C$
if one assumes that~$M$ is a finite set and~$\mu$ the counting measure.
We point out that the counter Example~\ref{ex29} works only for a continuous measure, but the
situation for discrete measures is unclear.
A partial result in this direction was obtained in~\cite[Theorem~2.9]{discrete}, where
an analog of Theorem~\ref{thm5} was proved without the constraint~$\T[F] \leq C$ for
$M$ a finite set and~$\mu$ the counting measure, assuming in addition the trace constraint~(C1)
(see also Theorem~\ref{thmcm} below and the corresponding statement for the
local correlation matrices in Section~\ref{sec32}).
But it is an open (and seemingly difficult) problem to prove the same for the identity constraint~(C2).
\item[(C)] Treating the constraint in the variational principle of Theorem~\ref{thm5}
with the Lagrange multiplier method, one finds that every minimizer~$F$ is a critical point
of the functional $\Sact - \kappa \T$ for a suitable Lagrange multiplier~$\kappa \in \R$.
It is not clear what the value of the Lagrange multiplier is, nor how it depends on the parameter~$C$.
What is the range of the Lagrange multipliers if~$C$ varies over~$\R^+$?
Of particular interest are the negative values of~$\kappa$, because this case cannot be handled
using Theorem~\ref{thm4}.
\item[(D)] Almost nothing is known about uniqueness. Clearly, we have the freedom to
perform isomorphisms of the measure space~$M$ as well as unitary transformations of~$\C^f$. Moreover,
we are free to change the function~$F$ as long as the moment measures remain unchanged.
But are the minimizers unique up to these obvious transformations, at least if~$\mu$ is
a non-atomic measure?
\item[(E)] In the setting of Section~\ref{sec15}, we saw examples of non-trivial minimizers. Do distributional minimizers as in Example~\ref{ex3} also exist if~$\beta>0$? Are the minimizers in
the setting of Lemma~\ref{lemma112} or Lemma~\ref{lemma113} smooth?
\item[(F)] It would be interesting to know more about the regularity of the minimizers in
the general setting of Chapter~\ref{sec2}.
Is every minimizing measure~$\rho$ supported on a set~$\K^f \subset \F$?
Under which conditions is the set~$\K^f$ a submanifold of~$\K$? If yes, in which situations
is~$\rho$ absolutely continuous with respect to the Lebesgue measure on~$\K$? When is the
corresponding Radon-Nikodym derivative a continuous or even smooth function on~$\K^f$?
\end{itemize}
Answering these questions goes beyond the scope of this paper. We now proceed to the
applications.

\section{Variational Principles in Indefinite Inner Product Spaces} \label{sec3}
Let us briefly outline the physical context in which causal variational principles arise
(for details see~\cite{PFP} or the review articles~\cite{rev, lrev}).
The Dirac wave functions in Minkowski space~$(M, \lbra .,. \lket)$
(or more generally on a Lorentzian manifold) have four complex components, and they are endowed
with an indefinite inner product~$\Sl \psi | \phi \Sr$ of signature~$(2,2)$ (this inner product is
usually written as~$\Sl \psi | \phi \Sr = \overline{\psi} \phi$ with~$\overline{\psi} = \psi^\dagger
\gamma^0$ the so-called adjoint spinor and $\gamma^0 = \text{diag}(1,1,-1,-1)$). Integrating this
inner product over space-time defines an inner product on the wave functions,
\beq \label{iprod}
\bra \psi | \phi \ket = \int_M \Sl \psi | \phi \Sr\: d^4x\:.
\eeq
A system of Dirac particles can be described by an operator~$P$ which is built up of
an ensemble of wave functions,
\beq \label{Psum}
P \phi = -\sum_{a} \bra \psi_a | \phi \ket \: \psi_a \:,
\eeq
where the index~$a$ runs over all the quantum numbers of the occupied states of the system.
The operator~$P$ is referred to as the fermionic projector or, more generally,
the {\em{fermionic operator}}. In~\cite{PFP} it was proposed
to formulate the physical equations in terms of a variational principle for the
fermionic operator in space-time. It is a main advantage of this approach that the metric,
causal and even topological structure of space-time does not enter the variational principle.
This makes it possible to formulate the physical equations on a set of points, referred to
as {\em{discrete space-time}}. As a consequence of a spontaneous symmetry breaking
effect~\cite{osymm}, the fermionic operator induces additional structures on
the space-time points, and there is some evidence that for systems involving many particles and
space-time points, these structures give rise to the usual topological and causal structure
of the space-time continuum~\cite{lrev}. Moreover, our variational principle can be analyzed
in Minkowski space in the so-called {\em{continuum limit}}. In view of these developments,
it is of interest to analyze the variational principle both in the discrete and continuous settings.
In order to treat these two cases in a unified setting, $M$ is best desribed by a general measure
space.

Analyzing variational principles in the general setting~\eqref{iprod} and~\eqref{Psum}
leads to two convergence problems: First, the space-time integral~\eqref{iprod} need not be
finite, and secondly the sum in~\eqref{Psum} might diverge.
In order to avoid these problems, we shall assume that~$M$ has {\em{finite volume}}.
Furthermore, we make the sum in~\eqref{Psum} finite by considering
only a {\em{finite number of particles}}. In this chapter we shall give a general existence proof
in finite volume and for a finite number of particles.
In Chapter~\ref{sec4}, we will prove existence for homogeneous systems
even in infinite volume for an infinite number of particles, assuming merely a momentum cutoff.

Before introducing the general setup and stating our results, we mention one parti\-cular difficulty
in analyzing variational principles in indefinite inner product spaces. The inner
product~\eqref{iprod} and also the action of any physically reasonable variational principle are
invariant under transformations of the wave functions of the form
\beq \label{gauge}
\psi(x) \rightarrow U(x) \,\psi(x) \qquad  \text{with $U(x) \in \U(2,2)$}\:,
\eeq
referred to as {\em{local gauge transformations}}. Since the group~$\U(2,2)$ is non-compact,
there is a large freedom to change the wave functions without affecting the physical action.

\subsection{Definitions and Statement of Results} \label{sec31}
Let~$(M, \mu)$ be a measure space of total volume~$\mu(M)=1$ and~$(V, \Sl .|. \Sr)$
a $2n$-dimensional complex vector space, endowed with a non-degenerate sesquilinear
form of signature~$(n,n)$ (for basic definitions see~\cite{GLR}). We consider the
vector space~$H=L^2(M, V, d\mu)$ of square-integrable functions from~$M$ to~$V$
and introduce on~$M$ the sesquilinear form
\beq \label{iproddef}
\bra \psi | \phi \ket = \int_M \Sl \psi(x) | \phi(x) \Sr \:d\mu(x)\:.
\eeq
Thus~$(H, \bra .|. \ket)$ is an indefinite inner product space. A function~$\psi \in H$
has~$2n$ components and is called a {\em{wave function}}. The parameter~$n$ is referred
to as the {\em{spin dimension}}. We introduce the {\em{fermionic operator}}~$P : H \rightarrow H$
as an operator having the following properties (A) and~(B1) or (B2):
\begin{itemize}
\item[(A)] $P$ is {\em{symmetric}} in the sense that
\[ \bra P \psi | \phi \ket = \bra \psi | P \phi \ket \qquad \text{for all~$\phi, \psi \in H$}\:. \]
\item[(B1)] The operator $P$ has {\em{finite rank}}, $\dim P(H) \leq f < \infty$.
It satisfies the {\em{trace constraint}} \label{pagetc}
\[ \tr(P)=f \]
(where~``$\tr$'' denotes the trace of operators in~$H$). Furthermore, the operator~$(-P)$ is
{\em{positive}} in the sense that
\[ \bra \psi | (-P) \psi \ket \geq 0 \qquad \text{for all $\psi \in \H$} \:. \]
\item[(B2)] The operator $P$ is a {\em{projector}} on a {\em{negative definite}} subspace of~$H$
of {\em{dimension}}~$f < \infty$.
\end{itemize}
In case~(B2), the operator~$P$ is called {\em{fermionic projector}}. In this case, the calculations
\[ \tr(P) = \text{rank}(P) = f\:,\qquad \bra \psi | (-P) \psi \ket = -\bra \psi | P^2 \psi \ket =
-\bra P \psi | P \psi \ket \geq 0 \]
show that condition~(B1) is again satisfied. Thus~(B2) is a special case of~(B1).
We refer to the rank of~$P$ as the {\em{number of particles}}. Note that for a fermionic
projector, the number of particles is equal to the parameter~$f$, whereas in case~(B1) the
number of particles is only bounded from above by~$f$.
For brevity, we refer to the system~$(P, H, \bra .|. \ket)$ as a {\em{fermion system}},
and~$M$ are the {\em{space-time points}}.

Since~$P$ has finite rank and~$\bra .|. \ket$ is non-degenerate, we can clearly
represent~$P$ by a finite matrix,
\[ P \psi = \sum_{k,l=1}^f \psi_k\, \bra \phi_l | \psi \ket \qquad \text{with $\psi_k, \phi_l \in H$}\:. \]
Writing the inner product with an integral~\eqref{iproddef}, one sees that~$P$ has an
integral representation,
\beq \label{intrep}
(P \psi)(x) = \int_M P(x,y)\, \psi(y)\: d\mu(y) \qquad \text{with} \qquad
P(x,y) \in L^2(M \times M, \Lin(V))
\eeq
(where~$\Lin(V)$ denotes the space of linear mappings from~$V$ to itself).
We refer to ~$P(x,y)$ as the {\em{kernel}} of the fermionic operator. Next, we introduce the
{\em{closed chain}}~$A_{xy}$ by
\beq \label{cchain}
A_{xy} = P(x,y)\, P(y,x) \:.
\eeq
Note that, for any fixed~$x$ and~$y$, the closed chain is a symmetric linear mapping of~$V$
to itself. More generally, for any endomorphism~$A$ of~$V$ we define
\beq \label{sweight}
\text{the spectral weight} \qquad |A| = \sum_{l=1}^{2n} |\lambda_l| \:,
\eeq
where~$\lambda_l$ are the eigenvalues of~$A$, counted with algebraic multiplicities.
We introduce
\beq \label{Lagdefi}
\boxed{ \quad \text{the Lagrangian} \quad\quad \L[A_{xy}] = |A_{xy}^2| - \frac{1}{2n}\: |A_{xy}|^2 
\quad}
\eeq
and define the functionals~$\Sact$ and~$\T$ by
\begin{align}
\Sact[P] &= \iint_{M \times M} \L[A_{xy}]\: d\mu(x)\, d\mu(y) \label{Sdefi} \\
\T[P] &= \iint_{M \times M} |A_{xy}|^2\: d\mu(x)\, d\mu(y) \label{Tdefi}\:.
\end{align}
These definitions are similar to~\eqref{Lagdef}--\eqref{Tdef}. However, we point
out that now the mathematical setting is very different; in particular the functionals now
depend on the fermionic operator~$P$.
Again using Definition~\ref{causal}, the eigenvalues~$\lambda^{xy}_j$ of the closed
chain~$A_{xy}$ induce on~$M$ a {\em{causal structure}}. Now we even get
a connection to the usual notion of causality: If the fermionic operator describes vacuum Dirac seas,
the causal structure of Definition~\ref{causal} coincides precisely with the causal structure of
Minkowski space (for details see~\cite[Section~6]{lrev}).

We prove the following existence results.
\begin{Thm} \label{thm4i} For given parameters~$f, n \in \N$ with~$f \geq 2n$
and a parameter~$\nu$ in the range~\eqref{nucond},
we consider the variational principle
\[ \text{minimize $\T[P] + \nu\, \Sact[P]$} \]
within the class of all fermionic operators with the properties (A) and
either~(B1) or~(B2). Then the minimum is attained.
\end{Thm}

\begin{Thm} \label{thm5i} For given parameters~$f, n \in \N$ with~$f \geq 2n$
and a constant~$C > 0$, we consider the variational principle
\[ \text{minimize $\Sact[P]$~on~${\mathcal{P}}_C := \{ P$ 
with $\T[P] \leq C \}$} \]
within the class of all fermionic operators with the properties (A) and
either~(B1) or~(B2). Then the minimum is attained.
\end{Thm} \noindent
The method of proof is to consider the corresponding local correlation matrices (see Section~\ref{sec23}),
making it possible to apply the results of Chapter~\ref{sec2}. In this reformulation, the
conditions~(B1) and~(B2) will correspond precisely to the conditions~(C1) and~(C2)
on page~\pageref{pagetc0}, respectively.

Before entering these constructions,
we briefly discuss the above theorems and put them in the context of previous work.
In~\cite{discrete} similar existence results were obtained in the special case where~$M$ is a
finite set and~$\mu$ the {\em{counting measure}}. These assumptions are a major simplification because
then the vector space~$H$ is finite-dimensional.
The remaining difficulty is to handle the non-compact gauge freedom~\eqref{gauge}, and this
problem is overcome by suitable a-priori estimates and a gauge fixing procedure.
In order to avoid confusion of notation, we point out that in~\cite{discrete} the Lagrangian~\eqref{Lagdefi}
is referred to as the critical Lagrangian. Also, instead of minimizing the functional
$\T[P] + \nu\, \Sact[P]$, in~\cite{discrete} the action corresponding to the
Lagrangian $\L_\mu = |A^2|-\mu |A|^2$ is considered for~$\mu > \frac{1}{2n}$.
These different functionals can easily be transformed into each other, and one sees that
the procedure in~\cite{discrete} is equivalent to minimizing $\T[P] + \nu\, \Sact[P]$
for any~$\nu > 0$. The procedure in the present paper has the advantage that we can also
consider negative values of~$\nu$.

There are a few general differences between the results in~\cite{discrete} and the
above theorems. First, in~\cite{discrete} the analog of Theorem~\ref{thm5i} is proved
under the constraint~$\T[P]=C$ (instead of~$\leq C$). This can be understood from the
fact that for a finite counting measure, the bubbling phenomenon of Section~\ref{sec23}
cannot appear (as is obvious from the inequality~\eqref{nobubble} for atoms of our measure).
Furthermore, it is worth noting that the following result for a finite counting measure
(see~\cite[Theorems~2.5 and~2.9]{discrete}) does {\em{not}} carry over to continuous measures.
\begin{Thm} \label{thmcm}
Suppose that~$\mu$ is a finite counting measure. Then the variational principle
\[ \text{minimize $\Sact[P]$} \]
attains its minimum if considered in one of the following two situations:
\begin{enumerate}
\item The fermionic operator satisfies~(A) and~(B1).
\item The fermionic operator satisfies~(A) and~(B2), and~$P$
is homogeneous in the sense that $\Tr(P(x,x)) = \Tr(P(y,y))$ for all $x, y \in M$.
\end{enumerate}
\end{Thm} \noindent
The three-dimensional Dirac sphere (Example~\ref{ex29}) shows that this theorem does not
hold for continuous measures, because there are divergent minimal sequences.
Thus there are situations where minimizers exist only due to the discreteness of space-time.

We finally point out that in~\cite{small} examples of minimizers are constructed
for a small number of particles and space-time points, and the resulting causal structure
is discussed.

\subsection{Reformulation in Terms of Local Correlation Matrices} \label{sec32}
A direct approach to proving Theorems~\ref{thm4i} and~\ref{thm5i} seems difficult
because the inner product space~$(H, \bra .|. \ket)$ is infinite-dimensional.
In such infinite-dimensional indefinite inner product spaces (also called Krein spaces,
see for example~\cite{bognar}), the functional analytic methods are quite limited,
making it hard to control the behavior of minimizing sequences~$(P_k)$ of
our variational principles. A particular problem in this setting
is the above-mentioned gauge freedom~\eqref{gauge}.
In order to bypass these difficulties, we shall proceed differently, making essential
use of the fact that the operator~$P$ has finite rank.
More precisely, our method is to choose convenient generators~$\psi_1, \ldots, \psi_f$
of the image of~$P$ and to consider the so-called {\em{local correlation matrices}}~$F_x$ defined
by
\beq \label{Fxjk}
(F_x)^j_k = - \Sl \psi_j(x) | \psi_k(x) \Sr \:.
\eeq
We shall see that the functionals~$\Sact$ and~$\T$ as well as the constraints~(B1) and~(B2)
can be reformulated purely in terms of the local correlation matrices. This remarkable fact
will make it possible to apply the results of Chapter~\ref{sec2}, giving us minimizing
local correlation matrices. In Section~\ref{sec33} we will conclude the proof by constructing
a corresponding fermionic operator.

We first specify which generators~$\psi_1, \ldots, \psi_f$ of~$P(H)$ we want to
use in~\eqref{Fxjk}.
\begin{Lemma} There are vectors~$\psi_1, \ldots, \psi_f \in H$ such that
\beq \label{Prep}
P \psi = -\sum_{l=1}^f \psi_l \bra \psi_l | \psi \ket \qquad \text{for all $\psi \in H$}\:.
\eeq
\end{Lemma}
\Proof It is convenient to work in~$V$ with a fixed pseudo-orthonormal basis~$(\e_i)$
where the inner product has the standard representation with a signature matrix~$S$,
\beq \label{Vrep}
\Sl u | v \Sr = \langle u, S v \rangle_{\C^{2n}} \qquad \text{where} \qquad
S = \text{diag}(\underbrace{1, \ldots, 1}_{\text{$n$ times}}, \underbrace{-1, \ldots, -1}_{\text{$n$ times}}) \:.
\eeq
The signature matrix can be regarded as an operator on~$V$. Alternatively, we
can consider~$S$ as an operator on~$H$, acting by pointwise multiplication
(i.e.\ $(S \psi)(x) := S (\psi(x))$). This makes it possible to represent the inner product on~$H$
in terms of the standard $L^2$ scalar product by
\beq \label{bkrep}
\bra \psi | \phi \ket = \lbra \psi, S \phi \lket_{L^2(M, V, d\mu)}\:.
\eeq
Obviously, $S^2=\1$ and $S$ is symmetric with respect to both~$\bra .|. \ket$ and~$\lbra .|. \lket_{L^2}$.

As the operator $(-P)$ is symmetric and positive
on~$(H, \bra .|. \ket)$, the operator~$(-SP)$ is symmetric and positive semi-definite
on the Hilbert space~$(H, \lbra .|. \lket_{L^2})$. Using furthermore that the operator~$(-SP)$
has rank at most~$f$, we can diagonalize this operator and choose orthonormal
eigenvectors $\phi_1, \ldots, \phi_f$ which span its image.
Since~$(-SP)$ is positive semi-definite, the corresponding eigenvalues $\kappa_1, \ldots, \kappa_f$
are non-negative. Thus the operator~$(-SP)$ has the representation
\[ (-SP)(\psi) = \sum_{l=1}^f \kappa_l\: \phi_l \lbra \phi_l | \psi \lket_{L^2} \:. \]
Multiplying by~$(-S)$ and using~\eqref{bkrep} together with the fact that~$S^2=\1$ and that~$S$
symmetric with respect to~$\lbra .|. \lket_{L^2}$, we obtain a similar representation for~$P$,
\[ P \psi = -\sum_{l=1}^f \kappa_l\: S \phi_l \bra S \phi_l | \psi \ket\:. \]
Setting~$\psi_l = \sqrt{\kappa_l}\: S \phi_l$ the result follows.
\QED

We next rewrite the fermionic operator and the local correlation matrices in a compact form.
Comparing~\eqref{Prep} with~\eqref{intrep} and~\eqref{iproddef}, we find that
\beq \label{Pxyv}
P(x,y) v = -\sum_{l=1}^f \psi_l(x) \Sl \psi_l | v \Sr \qquad \text{for all $v \in V$}\:.
\eeq
Introducing the operators
\beq \left. \begin{split}
\iota_x &\::\: V \rightarrow \C^f \::\: v \mapsto \left(\Sl \psi_l(x) | v \Sr \right)_{l=1,\ldots, f} \\
e_x &\::\: \C^f \rightarrow V \::\: u \mapsto \sum_{l=1}^f u_l \, \psi_l(x)\:, 
\end{split} \qquad \right\} \label{pidef}
\eeq
the identities~\eqref{Pxyv} and~\eqref{Fxjk} can be written in the simple form
\beq \label{Pkerrep}
\boxed{ \quad 
\begin{split} P(x,y) & = -e_x \,\iota_y \,:\, V \rightarrow V \\
F_x &= -\iota_x \,e_x \,:\, \C^f \rightarrow \C^f\:. \end{split} \quad }
\eeq

The next lemma gives the connection to the setting of Chapter~\ref{sec2}.
\begin{Lemma}
\begin{itemize}
\item[(i)] For every~$x \in M$, the local correlation matrix~$F_x$ is Hermitian. It has at most~$n$
positive and at most~$n$ negative eigenvalues.
\item[(ii)] In cases~(B1) or~(B2) (on page~\pageref{pagetc}), the function~$F(x)=F_x$
satisfies the conditions~(C1) and~(C2) (on page~\pageref{pagetc0}), respectively.
\item[(iii)] Denoting the eigenvalues of the matrix~$F_x F_y$ as in~\eqref{lambdacount},
the~$\lambda^{xy}_1, \ldots, \lambda^{xy}_{2n}$
coincide precisely with the eigenvalues of the closed chain~$P(x,y)\, P(y,x)$, counted with
algebraic multiplicities.
\end{itemize}
\end{Lemma}
\Proof It is obvious from the definition~\eqref{Fxjk} that $F_x$ is Hermitian.
Also, the number of its positive and negative eigenvalues is bounded by the signature~$n$ of~$V$.
This proves~(i).

To prove~(ii), we integrate~\eqref{Fxjk} over~$x$ and use~\eqref{iproddef} to obtain
\beq \label{Fxint}
\int_M (F_x)^j_k \,d\mu(x) = -\bra \psi_j | \psi_k \ket\:.
\eeq
Taking the trace and using~\eqref{Prep}, we conclude that condition~(B1) implies~(C1).
If~(B2) is satisfied, the rank of~$P$ equals~$f$, and thus the vectors~$\psi_1, \ldots, \psi_f$
in~\eqref{Prep} are linearly independent. Taking the square of~\eqref{Prep} and using
the idempotence of~$P$, we obtain another representation of~$P$,
\[ P \psi = \sum_{k,l} \psi_k \bra \psi_k | \psi_l \ket \bra \psi_l | \psi \ket \:. \]
Comparing these two representations in the basis~$\psi_1, \ldots, \psi_f$ of~$P(H)$, it follows
that
\[ \bra \psi_j | \psi_k \ket = -\delta_{jk} \:. \]
In view of~\eqref{Fxint}, we conclude that the identity constraint~(C2) holds.

To prove~(iii), we first recall that for quadratic matrices~$A$ and~$B$, the
products~$AB$ and~$BA$ have the same characteristic polynomials
(see for example~\cite[Section~3]{discrete}). If the matrices are not quadratic, i.e.\
$A \in \Mat(\C^p, \C^q)$, $B \in \Mat(\C^q, \C^p)$ with~$p>q$, we can extend the
matrices by zero rows and columns to obtain quadratic~$p \times p$-matrices.
Using the above result for quadratic matrices, it follows that
\beq \label{ABBA}
\det(AB - \lambda \1_{\C^p}) = \lambda^{p-q} \:\det(BA - \lambda \1_{\C^q}) \:.
\eeq
In words, the matrices~$AB$ and~$BA$ have the same eigenvalues and algebraic multiplicities,
up to the zero eigenvalues whose multiplicity is obvious counting dimensions.

We now write the product of the local correlation matrices according to~\eqref{Pkerrep} as
\[ F_x F_y = A B \qquad \text{where} \qquad A = e_x \text{ and } B = \iota_x e_y \iota_y \:. \]
Applying~\eqref{ABBA}, we see that this matrix has (up to the obvious zero eigenvalues)
the same eigenvalues and multiplicities as the matrix
\[ B A = \iota_x e_y \iota_y e_x = P(x,y)\, P(y,x) \:, \]
where in the last step we used~\eqref{Pkerrep}. This proves~(iii).
\QED

This lemma shows that considering the function~$F(x)=F_x$, we are precisely
in the setting of Chapter~\ref{sec2}. In particular, Theorems~\ref{thm4} and~\ref{thm5} 
yield minimizing local correlation matrices.

\subsection{Reconstruction of the Fermion System} \label{sec33}
The next lemma shows that the local correlation matrices can always be represented by
a suitable fermionic operator~$P$.

\begin{Lemma}
For any function~$F \in L^2(M, \F, d\mu)$ there is a kernel~$P \in L^4(M \times M, \Lin(V))$
such that for every~$x \in M$, the local correlation matrix~\eqref{Pkerrep} coincides with~$F(x)$.
The corresponding fermionic operator~$P$ satisfies the condition~(A).
If the conditions~(C1) or~(C2) (on page~\pageref{pagetc0}) are satisfied, then the fermionic operator
satisfies the conditions~(B1) or~(B2) (on page~\pageref{pagetc}), respectively.
\end{Lemma}
\Proof For given~$x \in M$ we diagonalize the matrix~$F(x)$ by a unitary transformation~$U$,
\[ F(x) = U D U^{-1} \quad \text{with} \quad U \in \U(\C^f) \text{ and }
D = \text{diag}(\nu_1, \ldots, \nu_f)\:. \]
We order the eigenvalues such that~\eqref{nudenote} holds and~$\nu_{2n+1}, \ldots, \nu_f = 0$.
Hence introducing the matrices
\beq \label{rhosdef}
\rho = \text{diag}(\sqrt{|\nu_1|}, \ldots, \sqrt{|\nu_f|}) \quad \text{and} \quad
s = \text{diag}(\underbrace{1, \ldots, 1}_{\text{$n$ times}},
\underbrace{-1, \ldots, -1}_{\text{$n$ times}},
\underbrace{0, \ldots, 0}_{\text{$f-2n$ times}} ) \:,
\eeq
we have the decomposition
\[ F(x) = - U \rho s \rho U^{-1}\:. \]
We now introduce the vectors~$\psi_1(x), \ldots, \psi_k(x) \in V$ by
\[ (\psi_l)^a = (\rho U^{-1})^a_l  \:,\qquad a=1,\ldots, 2n\:, \]
where in~$V$ we again work in the basis where the inner product has the form~\eqref{Vrep}.
Carrying out this construction for every~$x \in M$, we obtain functions~$\psi_l \,:\,
M \rightarrow V$. Since~$F \in L^2$, its eigenvalues~$\nu_j$ are square integrable,
and due to the square roots in~\eqref{rhosdef} and the unitarity of the transformation~$U$,
the functions~$\psi_l$ are in $L^4(M, V, d\mu) \subset H$. Hence~\eqref{Prep} defines a fermionic
operator~$P$ with~$P(x,y) \in L^4(M \times M, \Lin(V), d\mu)$. Furthermore, from the
above construction it follows that
\beq \label{Fxkl2}
(F_x)^k_l = -\langle \psi_k(x), S \psi_l(x) \rangle_{\C^{2n}} = -\Sl \psi_k(x), \psi_l(x) \Sr \:,
\eeq
and thus in view of~\eqref{Fxjk} we see that the local correlation matrix~$F_x$ indeed
coincides with the matrix~$F(x)$.

Taking the trace of~\eqref{Fxint} and using~\eqref{Prep} we obtain
\[ \tr P = - \int_M \Tr(F(x)) \: d\mu(x)\:, \]
and thus~(C1) indeed implies~(B1). If~(C2) holds, we obtain from~\eqref{Fxint} that
the vectors~$\psi_1, \ldots, \psi_f$ are mutually orthogonal and normalized to~$\bra \psi_l | \psi_l \ket=-1$.
Hence the fermionic operator~\eqref{Prep} is a projector on a negative definite subspace of~$H$
of dimension~$f$.
\QED

In view of this lemma, Theorem~\ref{thm4i} and Theorem~\ref{thm5i} follow immediately
from Theorem~\ref{thm4} and Theorem~\ref{thm5}, respectively.

\section{A Variational Principle in Infinite Space-Time Volume} \label{sec4}
As explained at the beginning of Chapter~\ref{sec3}, the variational principles so far
were restricted in that the space-time volume and the number of particles had to be finite.
We now introduce a class of variational principles in {\em{infinite space-time volume}},
which may involve an {\em{infinite number of particles}}. On the other hand,
we need to specialize our setting by assuming that our
system is {\em{homogeneous}} in the sense that the
kernel of the fermionic projector~$P(x,y)$ depends only on the difference vector~$y-x$.
This makes it necessary to assume that space-time has an underlying vector space structure,
and thus this setting is of main interest in two situations: in the so-called {\em{continuous case}}
we assume that~$(M, \lbra .,. \lket)$ is Minkowski space and~$\mu$ is the Lebesgue measure on~$M$, whereas in the so-called {\em{discrete case}} we assume that~$(M, \lbra .,. \lket)$ is a
periodic lattice in Minkowski space (i.e.\ a discrete subgroup of~$\R^4$) and~$\mu$ is the
counting measure.
Generalizations of these two cases are clearly possible, but will not be considered here.

The main simplification in the homogeneous setting is that~$P(x,y)$ can be written as a
Fourier transform, i.e.\ in the usual physics notation
\[ P(x,y) = \int_{\hat{M}} \frac{d^4p}{(2 \pi)^4} \: \hat{P}(p)\: e^{i\! \lbra p, y-x \lket} \:, \]
where~$\hat{M}$ is momentum space, which in the continuum case is isomorphic to
Minkowski space, whereas in the discrete case~$\hat{M}$ is a primitive cell of the reciprocal lattice.
In order to obtain a nice measure-theoretic framework, we here combine the product $\hat{P}(p) \,d^4 p$
into a Borel measure~$d\nu$, taking values in~$\Lin(V)$, where~$(V, \Sl .|. \Sr)$ is again an
indefinite inner product space of signature~$(n,n)$. In order to describe a completely filled Dirac sea
in the vacuum, the measure~$d\nu$ would take the form (see for example~\cite[Section~6]{lrev})
\beq \label{nusea}
d\nu(p) = (p_j \gamma^j + m)\: \delta(\lbra p,p \lket - m^2)\: \Theta(-p^0)\: d^4p\:,
\eeq
where~$\gamma^j$ are the Dirac matrices in Minkowski space, and~$V=\C^4$ endowed
with the inner product $\Sl \psi | \phi \Sr = \overline{\psi} \phi$, where~$\overline{\psi}=\psi^\dagger \gamma^0$ is the adjoint spinor. This measure has the
remarkable property that $-\nu$ is positive in the sense that
\beq \label{nup}
\Sl v | (-\nu(\Omega)) v \Sr \geq 0 \qquad \text{for all $v \in V$}\:.
\eeq
Unfortunately, measures of the form~\eqref{nusea} lead to ultraviolet problems.
In order to avoid these problems, we shall assume that the measure~$d\nu$ is supported
in a given bounded subset~$\hat{K} \subset \hat{M}$ (this is clearly no restriction in the
discrete case, where already~$\hat{M}$ is bounded). This motivates the following definition.

\begin{Def} \label{defnegV}
Consider a regular Borel measure~$\nu$ on a bounded set~$\hat{K} \subset \hat{M}$ taking
values in~$\Lin(V)$ with the following properties:
\begin{itemize}
\item[(i)] For every~$v \in V$, the measure~$d \Sl v | \nu v \Sr$ is a finite real measure.
\item[(ii)] For every Borel set~$\Omega \subset \hat{K}$, the operator~$-\nu(\Omega)
\in \Lin(V)$ is positive~\eqref{nup}.
\end{itemize}
Then~$\nu$ is called a {\bf{negative definite measure}} on~$\hat{K}$ with values in~$\Lin(V)$.
\end{Def} \noindent
For a given negative definite measure~$\nu$, we introduce the {\em{kernel}} of the fermionic operator by
\beq \label{Pxidef}
P(\xi) = \int_{\hat{K}} e^{i\!\lbra p, \xi \lket}\: d\nu(p)\:,\qquad \text{where $\xi \equiv y-x$}\:.
\eeq
Introducing the corresponding {\em{fermionic operator}}~$P$ by
\[ (P \psi)(y) = \int_M P(y-x)\: \psi(x)\: d\mu(x)\:, \]
this operator is well-defined for example on the test functions~$C^\infty_0(M, V)$.
We point out that this operator will in general have infinite
rank, and thus (adopting the notion from Section~\ref{sec31})
the total the number of particles is infinite. However, the quantity
\beq \label{locdens}
f_{\text{loc}} := \Tr (P(0))
\eeq
is finite; it can be interpreted as the local particle density, which for our homogeneous system
is constant in space-time.

Following the procedure in Chapter~\ref{sec3}, we introduce the closed
chain by~$A(\xi) = P(\xi) P(-\xi)$ and define the spectral weight and the Lagrangian
by~\eqref{sweight} and~\eqref{Lagdefi}, respectively. However, the space-time integrals in~\eqref{Sdefi}
and~\eqref{Tdefi} do not converge, because rewriting them as
\[ \iint_{M \times M} d\mu(x) \:d\mu(y) \cdots = \int_{M} d\mu(x) \int_{M} d\mu(\xi) \cdots \]
and carrying out the~$\xi$-integral, our homogeneity assumption implies that the resulting function
is constant, so that the~$x$-integral necessarily diverges.
This divergence is an artifact of working in infinite volume. Keeping in mind that
in finite volume (obtained for example by compactifying space-time to a torus) the $x$-integral
would give rise to an irrelevant constant, it is natural to simply drop the outer integral.
This leads us to introduce the functionals~$\Sact$ and~$\T$ by
\beq \label{STnu}
\Sact[\nu] = \int_M \L[A(\xi)]\, d\mu(\xi) \:,\qquad
\T[\nu] = \int_M |A(\xi)|^2\, d\mu(\xi) \:.
\eeq

We are now ready to state the main result of this chapter.
\begin{Thm} \label{thminf} Let~$(\nu_k)_{k \in \N}$ be a sequence of negative definite measures
on the bounded set~$\hat{K} \subset \hat{M}$ satisfying one of the following two conditions:
\begin{itemize}
\item[(I)] Assume that~$(M, \mu)$ is Minkowski space with the Lebesgue measure or
a lattice with the counting measure. Assume furthermore that the functional~$\T$ is bounded,
\[ \T[\nu_k] \leq C \qquad \text{for all $k \in \N$}\:. \]
\item[(II)] Assume that~$(M, \mu)$ is a lattice with the counting measure.
Assume furthermore that the functional~$\Sact$ is bounded,
\[ \Sact[\nu_k] \leq C \qquad \text{for all $k \in \N$}\:, \]
and that the local particle density is bounded away from zero in the sense that there is a constant~$\varepsilon>0$
with
\beq \label{ltrace}
\Tr \!\left( P[\nu_k](0) \right) \geq \varepsilon \qquad \text{for all~$k \in \N$} \:.
\eeq
\end{itemize}
Then there is a subsequence~$(\nu_{k_l})$ and a series of unitary transformations~$U_l$ on~$V$
such that the measures $U_l \nu_{k_l} U_l^{-1}$ converge in the $C^0(\hat{K})^*$-topology
to a negative definite measure~$\nu$ with the properties
\[ \T[\nu] \leq \liminf_k \T[\nu_k] \:,\qquad
\Sact[\nu] \leq \liminf_k \Sact[\nu_k]\:. \]
\end{Thm} \noindent
We note that, in contrast to the previous existence theorems, this theorem is stated as a compactness
result. By applying it to a minimal sequence, one immediately obtains statements similar
to those in Theorems~\ref{thm4} and~\ref{thm5}. The advantage of the above compactness
statement is that it applies immediately to situations where additional constraints are imposed
(like further conditions on the support, the assumption of a vector-scalar structure,
the condition of half-occupied surface states, etc).
One only needs to ensure that these constraints are invariant under the unitary transformations~$U_l$,
and that they are continuous in the $C^0(\hat{K})^*$-topology. As a simple example, one may prescribe
the local particle density~\eqref{locdens}.

Before coming to the proof, we illustrate the statement of the above theorem by a counter example,
which shows why the support of the measures~$\nu_k$ must be uniformly bounded
and why in case~(I) one may not replace the functional~$\T$ by the action~$\Sact$.

\begin{Example} (The Dirac cylinder) \em We let~$(M, \lbra .|. \lket)$
be Minkowski space with~$\mu$ the Lebesgue measure.
We choose~$n=2$ and~$(V, \Sl .|. \Sr)$ as in the example~\eqref{nusea}.
For given parameters~$\tau, L>0$, we consider in momentum space~$p=(\omega, \vec{p})$
the measure
\[ d\nu(p) = \frac{1}{16 \pi}\: \frac{\Theta(L-|\omega|)}{L}\:
\delta \!\left( |\vec{p}|^2-1\right) \left[ -\sqrt{\tau^2 + 1}\: \gamma^0 + 
\tau \vec{p} \vec{\gamma} + \1 \right] d^4p \:. \]
This measure is supported on the cylinder~$[-L,L] \times S^2$. Writing the square bracket
as~$(l^j \gamma_j + \1)$, the vector field~$l$ has Lorentz length one. Increasing~$\tau$
describes a Lorentz boost of this vector field.

As is easily verified, $\nu$ is a negative definite measure.
Computing its Fourier transform, we obtain
\begin{align*}
P(\xi) &= \int_{\hat{M}} e^{i \! \lbra p, \xi \lket}\: d\nu(p) \\
&= \frac{1}{8 \pi} \: \frac{\sin L t}{L t}
\left( -\sqrt{\tau^2 + 1}\: \gamma^0 + i \tau \vec{\gamma} \vec{\nabla} + \1 \right)
\int_{\R^3} \delta \!\left(|\vec{p}|^2-1\right) e^{-i \vec{p} \vec{\xi}} d\vec{p} \\
&= \frac{1}{4} \: \frac{\sin L t}{L t}
\left( -\sqrt{\tau^2 + 1}\: \gamma^0 + i \tau \vec{\gamma} \vec{\nabla} + \1 \right)
\: \frac{\sin |\vec{\xi}|}{|\vec{\xi}|} \\
&= \frac{1}{4} \: \frac{\sin L t}{L t} \left\{
\left( -\sqrt{\tau^2 + 1}\: \gamma^0 + \1 \right)
\: \frac{\sin r}{r} \:+\:\frac{i \tau \vec{\xi} \vec{\gamma}}{r^2}
\left( \cos r - \frac{\sin r}{r} \right) \right\} ,
\end{align*}
where in the last line we set~$r=|\vec{\xi}|$. In particular,
\[ \Tr(P(0)) = \frac{1}{4}\: \Tr \!\left( -\sqrt{\tau^2 + 1}\: \gamma^0 + \1 \right) = 1 \:, \]
so that the local particle density~\eqref{ltrace} is fixed. A straightforward calculation yields
\beq \label{Texp}
\T = \frac{\pi^3}{90\: L} \left( 3 \tau^4 + 10 \tau^2 + 15 \right) \:,
\eeq
and this diverges as~$\tau \rightarrow \infty$.
The causal structure is more interesting because for large~$\tau$, the timelike region shrinks
to a small cylinder~$r \leq r_{\max}$. More precisely, a Taylor expansion near the
origin yields for the two distinct eigenvalues~$\lambda_\pm$ of~$A(\xi)$
\[ (\lambda_+ - \lambda_-)^2 =  16 \left( \frac{1}{4} \: \frac{\sin L t}{L t} \right)^4 
(1+\tau^2) \left(1 - \frac{r^2 (6+\tau^2)}{9} \right) + {\mathscr{O}}(r^{-4})\:.  \]
Hence the timelike region is given by
\[ r < r_{\max} := \frac{3}{\tau} + {\mathscr{O}}\!\left( \frac{1}{\tau^3} \right) . \]
After verifying that in this region, the two eigenvalues~$\lambda_+$ and~$\lambda_-$ have the same
sign, the action is computed to be
\beq \label{Sexp}
\Sact = \frac{3 \pi^2}{5} \: \frac{1}{L \tau} + {\mathscr{O}}\!\left( \frac{1}{\tau^3} \right) .
\eeq
Hence the action tends to zero as~$\tau \rightarrow \infty$. We conclude
that~$\nu_k := \nu|_{\tau=k}$ is a minimizing sequence of the action~$\Sact$
which does not converge.

It is also worth noting that, according to~\eqref{Sexp} and~\eqref{Texp},
both functionals~$\Sact$ and~$\T$ become small if~$L$ tends to infinity.
Hence dropping the condition $\text{supp} \,\nu_k \subset \hat{K}$, the
sequence~$\nu_k := \nu|_{L=k}$ is a minimizing sequence which converges
in the $C^0(\hat{M})^*$-topology to the trivial measure~$\nu=0$.
\QEDrem \end{Example}

A possible method for proving Theorem~\ref{thminf} would be to proceed as in
Section~\ref{sec32} by choosing generators of $P(H)$ and considering the corresponding
local correlation matrices. However, since the rank of~$P$ may be infinite, it now seems easier to
analyze the kernel of the fermionic operator~$P(\xi)$ as an operator on the finite-dimensional 
indefinite inner product space~$(V, \Sl .|. \Sr)$.
This makes it possible to apply similar methods as in the previous chapters, which will be
complemented by suitable estimates for negative definite measures.
On~$V$ we again fix the pseudo-orthonormal basis~$(\e_i)$ in which the inner product~$\Sl .|. \Sr$
has the representation~\eqref{Vrep}.
Furthermore, we denote the sup-norm of matrices in this basis by~$\|.\|$ and consider it as a norm
on~$\Lin(V)$. We say that a linear operator~$B$ on~$V$ is {\em{positive}} if
\[ \Sl u | B u \Sr \geq 0 \quad \text{for all $u \in V$}\:. \]
Clearly, every positive operator is symmetric in~$V$. However, in contrast to the situation in
scalar product spaces, not every positive operator is diagonalizable, as one sees in the simple
two-dimensional example
\[ B = \begin{pmatrix} 1 & 1 \\ -1 & -1 \end{pmatrix}\:,\qquad
S = \begin{pmatrix} 1 & 0 \\ 0 & -1 \end{pmatrix} . \]
But a positive operator can be diagonalized up to an arbitrarily small error term,
as the next lemma shows.
\begin{Lemma} \label{lemma44} Suppose that~$B$ is a positive linear operator on~$V$. Then for every~$\varepsilon>0$
there is a unitary transformation~$U$ on~$(V, \Sl .|. \Sr)$ such that
\[ U B U^{-1} = -\text{\rm{diag}}(\nu_1, \ldots, \nu_{2n}) + \Delta B \:, \]
where the real parameters~$\nu_i$ are ordered as in~\eqref{nudenote} and
\[  \|\Delta B\|< \varepsilon \:. \]
\end{Lemma}
\Proof Suppose that the characteristic polynomial of~$B$ has a root~$\nu \neq 0$.
Then there is a corresponding eigenvector~$u \neq 0$. Since~$(V, \Sl .|. \Sr)$ is non-degenerate,
there is a vector~$v \in V$ with~$\Sl v | u \Sr \neq 0$. Then
\[ \Sl v | B u \Sr = \nu \Sl v | u \Sr \neq 0\:. \]
Since~$B$ is positive, the bilinear form~$\Sl .| B . \Sr$ is positive semi-definite.
The corresponding Schwarz inequality (see~\cite[Lemma~4.1~(ii)]{discrete} for details)
\beq \label{schwarz}
\Sl v | B u \Sr^2 \leq \Sl u | B u \Sr\, \Sl v | B v \Sr
\eeq
implies that
\[ 0 < \Sl u | B u \Sr = \nu \Sl u | u \Sr \:. \]
Hence~$u$ cannot be a neutral vector. Furthermore, the terms~$\nu$ and~$\Sl u | u \Sr$ have the
same sign. Introducing the pseudo-normalized vector~$f = u/\sqrt{|\Sl u|u \Sr|}$, the
orthogonal complement of~$f$ is an invariant subspace
which does not contain~$u$. Hence proceeding inductively, we can diagonalize~$B$
except on the invariant subspace corresponding to~$\nu=0$.

Restricting attention to this remaining invariant subspace, the characteristic polynomial of~$B$ is trivial. Then~$B$ is nilpotent, and thus we can choose a basis where~$B$ is a direct sum of Jordan chains.
Labeling the Jordan chains by an index~$c$ and the basis within each chain
by~$\f^{(c)}_1, \ldots, \f^{(c)}_{L(c)}$, we have
\[ B \f^{(c)}_1 = 0 \qquad \text{and} \qquad
B \f^{(c)}_{l+1} = \f^{(c)}_l \:, \quad l \in \{1, \ldots, L(c)-1\}\:. \]
Let us verify that the inner products between the basis vectors have the properties that
\beq \label{chainc}
\Sl \f^{(c)}_l | \f^{(c')}_{l'} \Sr = 0 \quad \text{unless $L(c)=L(c')$ and $l'+l=L(c)+1$} \:.
\eeq
To this end, we let~$c$ label the longest Jordan chain. Choosing an index $l'<L(c')$,
the symmetry of the operator~$B$ yields that
\[ \Sl \f^{(c)}_1 | \f^{(c')}_{l'}\Sr = \Sl \f^{(c)}_1 | B \f^{(c')}_{l'+1}\Sr
=  \Sl B \f^{(c)}_1 | \f^{(c')}_{l'+1}\Sr = 0 \:. \]
Hence the inner product~$\Sl \f^{(c)}_1 | \f^{(c')}_{l'}\Sr$ vanishes unless~$l'=L(c')$.
Next, from the calculation
\[ \Sl \f^{(c)}_1 | \f^{(c')}_{L(c')} \Sr = \Sl B^{L(c)} \f^{(c)}_{L(c)} | \f^{(c')}_{L(c')} \Sr
= \Sl \f^{(c)}_{L(c)} | B^{L(c)} \f^{(c')}_{L(c')} \Sr = 0 \quad \text{if $L(c') < L(c)$} \]
and the fact that~$c$ is the longest Jordan chain, we conclude that
\[ \Sl \f^{(c)}_1 | \f^{(c')}_{l'} \Sr = 0 \qquad \text{unless~$l' = L(c') = L(c)=:L$}\:. \]
Since~$\Sl .|. \Sr$ is non-degenerate, we know that there is a Jordan chain~$c'$ such
that the inner product~$\Sl \f^{(c)}_1 | \f^{(c')}_{L(c')} \Sr$ is non-zero. If possible, we choose~$c'=c$.
Then all the inner products between the basis vectors of the chains~$c$ and~$c'$ can be
computed as follows,
\[ \Sl \f^{(c)}_l | \f^{(c')}_{l'} \Sr = \Sl \f^{(c)}_l | B^{L-l'} \f^{(c')}_{L} \Sr 
=  \Sl B^{L-l'} \f^{(c)}_l | \f^{(c')}_{L} \Sr \\
=  \Sl \f^{(c)}_{l+l'-L} | \f^{(c')}_{L} \Sr \:. \]
The last inner product clearly vanishes if~$l+l'-L<1$, whereas it is non-zero if~$l+l'-L=1$.
In the remaining case~$l-l'-L>1$, we can make the inner product to zero by the transformation
\[ \f^{(c)}_{l+l'-L} \rightarrow \f^{(c)}_{l+l'-L} - \frac{\Sl \f^{(c)}_{l+l'-L} | \f^{(c')}_{L} \Sr}
{\Sl \f^{(c)}_{1} | \f^{(c')}_{L} \Sr}\: \f^{(c)}_{1} \]
(note that the vector~$\f^{(c)}_{1}$ is in the kernel of~$B$, and thus the matrix elements of~$B$
remain unchanged). After these transformations, the relations~\eqref{chainc} are
satisfied for the chains~$c$ and~$c'$. Furthermore, one easily verifies that the subspace of~$V$ spanned by
the vectors~$\f^{(c)}_l$ and~$\f^{c'}_l$ is non-degenerate. Hence by going over to its
orthogonal complement, we can proceed iteratively, proving~\eqref{chainc}.

According to~\eqref{chainc}, all the inner products between the basis vectors remain unchanged
if we rescale the basis vectors for any parameter~$\rho>0$ according to
\[ \f^{c}_l \rightarrow \rho^{\frac{L(c)+1}{2}-l} \: \f^{c}_l\:. \]
But this transformation multiplies all matrix entries of~$B$ by a factor~$\rho^{-1}$. Thus by
choosing~$\rho$ sufficiently large, we can arrange that the matrix entries of~$B$ become
arbitrarily small. We point out that the Jordan basis is not pseudo-orthonormal. But since
the transformation matrix from the Jordan basis to the pseudo-orthonormal basis can be
chosen independent of~$\rho$, we can make the matrix entries of~$B$ arbitrarily small
even in a pseudo-orthonormal basis.

We have shown that there is a pseudo-orthonormal basis~$(\f_i)$ in which~$B$ is diagonal up to
an arbitrarily small correction~$\Delta B$. Denoting the unitary transformation from the
basis~$(\f_i)$ to our original pseudo-orthonormal basis~$(\e_i)$ by~$U$
(i.e.\ $U(\f_i)=\e_i$), the result follows.
\QED

We next bound the Lagrangian at~$\xi=0$ from below, using a method
similar as in~\cite[Proposition~4.3]{discrete}.
\begin{Lemma} \label{lemma45}
For every negative definite measure~$\mu$, the corresponding Lagrangian
\eqref{STnu} satisfies at~$\xi=0$ the inequality
\[ \L[\nu](0) \geq \frac{1}{8 n^5}\: |P(0)|^2 \Tr(P(0))^2 \:, \]
where~$|.|$ again denotes the spectral weight.
\end{Lemma}
\Proof According to Definition~\ref{defnegV}~(ii), the operator
$(-P(0)) = -\nu(\hat{K})$ is positive. Thus applying Lemma~\ref{lemma44}
and taking the limit~$\varepsilon \searrow 0$,
we conclude that the spectrum of~$P$ is real, and that we can order its
eigenvalues~$\nu_i$ counted with algebraic multiplicities as in~\eqref{nudenote}.
The eigenvalues of~$A(0)=P(0)^2$ are then obviously given by~$\lambda_i=\nu_i^2 \geq 0$.
Hence rewriting the Lagrangian~\eqref{Lagdefi} according to~\eqref{Lcrit}
and choosing indices~$a, b \in \{1, \ldots, 2n\}$ such that~$|\nu_a|$ and~$|\nu_a|-|\nu_b|$ are maximal,
we obtain the estimates
\begin{align*}
\L[A(0)] &= \frac{1}{4n} \sum_{i,j=1}^{2n} (\lambda_i-\lambda_j)^2
= \frac{1}{4n} \sum_{i,j=1}^{2n} (|\nu_i|+|\nu_j|)^2 \:(|\nu_i|-|\nu_j|)^2  \\
& \geq \frac{1}{2n} \: |\nu_a|^2 \: (|\nu_a|-|\nu_b|)^2 
\geq \frac{1}{8n^3} \: |P(0)|^2 \: (|\nu_a|-|\nu_b|)^2 \\
\Tr(P(0)) &= \sum_{k=n+1}^{2n} |\nu_k| - \sum_{l=1}^n |\nu_l|
= \frac{1}{n} \sum_{k=n+1}^{2n} \sum_{l=1}^n \left( |\nu_k| - |\nu_l| \right)
\leq n\:(|\nu_a|-|\nu_b|)\:.
\end{align*}
Combining these inequalities gives the result.
\QED

\Proof[Proof of Theorem~\ref{thminf}]
For ease in notation, we omit the brackets~$[\nu]$ and replace the arguments~$[\nu_k]$ by an index,
i.e.\ $P_k \equiv P[\nu_k]$, $A_k \equiv A[\nu_k]$, $\Sact \equiv \Sact[\nu]$, and so on.
For any~$\varepsilon>0$ and~$k \in \N$,
we apply Lemma~\ref{lemma44} to the operator~$B=-\nu_k(\hat{K})$,
which is positive according to~\eqref{nup}. Replacing the measures~$\nu_k$
by~$U \nu_k U^{-1}$, we conclude that
\beq \label{nukrep}
\nu_k(\hat{K}) = \text{\rm{diag}}(\nu^{(k)}_1, \ldots, \nu^{(k)}_{2n}) + \Delta \nu_k 
\quad \text{with} \quad \| \Delta \nu_k \| \leq \varepsilon
\eeq
(where we again work in the pseudo-orthonormal basis~$(\e_i)$).
Our main task is to show that the measures~$\nu_k$ are uniformly bounded in the sense that 
there is a constant~$C>0$ such that
\beq \label{unibound}
\| \nu_k(\Omega) \| \leq C \qquad \text{for all $k \in \N$ and all Borel sets $\Omega \subset \hat{K}$} \:.
\eeq
We first complete the proof of the theorem assuming this uniform bound,
which we shall prove afterwards. If~\eqref{unibound} holds,
the Banach-Alaoglu theorem and the Riesz representation theorem
yield that a subsequence of the~$\nu_k$ converges in the $C^0(\hat{K})^*$-topology
to a negative definite measure~$\nu$. Thus the integral~\eqref{Pxidef} converges pointwise,
\[ P_k(\xi) \rightarrow P(\xi) \qquad \text{for every~$\xi \in M$}\:. \]
As a consequence, the spectral weight~$|A_k(\xi)|$ converges pointwise.
Using that the spectral weight is non-negative, we can apply Fatou's lemma to obtain
\[ \T = \int_M \lim_{k \rightarrow \infty} |A_k(\xi)|^2\, d^4 \xi \leq
\lim_{k \rightarrow \infty} \int_M |A_k(\xi)|^2\, d^4 \xi = 
\lim_{k \rightarrow \infty} \T_k \:. \]
Similarly, one sees that~$\Sact \leq \lim_{k \rightarrow \infty} \Sact_k$.
This concludes the proof provided that~\eqref{unibound} holds.

To prove~\eqref{unibound}, we first make use of the positivity property~(ii)
in Definition~\ref{defnegV}. Namely, this property ensures that
for any Borel set~$\Omega \subset \hat{K}$, the bilinear form $\Sl .| (-\nu_k(\Omega)) . \Sr$ 
is positive semi-definite. Thus the Schwarz inequality yields
\[ \left( \Sl \e_i | (-\nu_k(\Omega)) \e_j \Sr \right)^2 \leq 
\Sl \e_i | (-\nu_k(\Omega)) \e_i \Sr\: \Sl \e_j | (-\nu_k(\Omega)) \e_j \Sr \:. \]
In other words, the off-diagonal diagonal matrix elements are bounded in terms of the
diagonal entries, and thus it suffices to show that the diagonal entries~$\Sl \e_i | (-\nu_k(\Omega)) \e_i \Sr$
are bounded. Furthermore, in view of the fact that the measures~$\Sl \e_i | (-\nu_k(.)) \e_i \Sr$
are all positive, it suffices to show that the total measure is bounded, i.e.
\[ | \Sl \e_i | \nu_k(\hat{K}) \e_i \Sr | \leq C \qquad \text{for all~$k \in \N$}. \]
Applying~\eqref{nukrep} and taking the limit~$\varepsilon \searrow 0$, 
this condition reduces to demanding that the spectral weight of~$\nu_k(\hat{K})$ be bounded,
\beq \label{contcond}
|\nu_k(\hat{K})| \leq C \qquad \text{for all~$k \in \N$}.
\eeq

In case~(II) when~$\mu$ is a counting measure, it is obvious that~$\L_k(0) \leq \Sact_k$.
Thus Lemma~\ref{lemma45} yields
\[ |P_k(0)|^2 \leq 8 n^5\: \frac{\Sact_k}{|\Tr(P_k(0))|^2} \:. \]
In view of~\eqref{ltrace} and the uniform boundedness of the action,
we thus have a uniform a-priori bound for the spectral weight of~$P_k(0)$.
Using that~$P_k(0)=\nu_k(\hat{K})$, we obtain~\eqref{contcond}.

In case~(I) and if~$\mu$ is the counting measure, the uniform boundedness of
the functional~$\T$ yields that the spectral weight~$|A_k(0)|$ is uniformly bounded.
Since~$A_k(0) = P_k(0)^2$, it follows that
the spectral weight of~$P_k(0)$ is also uniformly bounded, proving~\eqref{contcond}.
Thus it remains to consider case~(I) for~$\mu$ the Lebesgue measure.
This is the most difficult case, because we need
to bound the space-time integral~\eqref{STnu} from below by~$|P(0)|$, making it
necessary to estimate~$|A(\xi)|$ in a neighborhood of~$\xi=0$. We introduce
the quantity~$\Delta P(\xi) = P(\xi)-P(0)$. Using~\eqref{Pxidef}, we obtain
\beq \label{es1}
\|\Delta P_k(\xi)\| \leq \int_{\hat{K}}
\left| e^{i \! \lbra p, \xi \lket} - 1\right| d \|\nu_k(p)\| \leq
{\mbox{diam}}(\hat{K})\, \|\xi\|\: \|\nu_k\|(\hat{K}) \:,
\eeq
where~$\|\xi\|$ is the Euclidean norm on~$\R^4$, and the norm of the measure
is defined by
\[ \|\nu\|(\hat{K}) = \sum_{i,j=1}^{2n} |\Sl \e_i | \nu \e_j \Sr|(\hat{K}) \]
(and~$|\Sl \e_i | \nu \e_j \Sr|$ denotes the variation of a complex-valued measure,
see~\cite[Sections~28 and~29]{halmosmt}).
In order to estimate this norm, we first note that the Schwarz inequality~\eqref{schwarz}
allows us to estimate the off-diagonal elements~$|\Sl \e_i | \nu \e_j \Sr|(\hat{K})$ in terms
of the diagonal elements. For the diagonal elements, on the other hand, we can use
that $-\Sl \e_i | \nu \e_i \Sr$ is a positive measure. Hence
\[ \|\nu(\hat{F})\| \leq n \,|\nu(\hat{F})|\:. \]
Using this inequality in~\eqref{es1}, we conclude that
\[ \|\Delta P_k(\xi)\| \leq \int_{\hat{K}}
\left| e^{i \! \lbra p, \xi \lket} - 1\right| d \|\nu_k(p)\| \leq
n\:{\mbox{diam}}(\hat{K})\, \|\xi\|\: |\nu_k(\hat{K})| \:. \]

We choose~$\xi$ so small that
\beq \label{xireg}
\| \xi\| \leq \frac{1}{8 n\, {\mbox{diam}}(\hat{K})} \:.
\eeq
Then~$P_k(\xi)$ has the representation
\[ P_k(\xi) = P_k(0) + \Delta P_k(\xi) \qquad \text{with} \qquad \|\Delta P_k(\xi)\| \leq \frac{1}{8}\:
|P_k(0)|\:. \]
Multiplying by a similar representation for~$P_k(-\xi)$ and
using that, according to~\eqref{nukrep}, the matrix~$P_k(0)$ is diagonal up to
an arbitrarily small error term, we conclude that
\[ A_k(\xi) = A_k(0) + \Delta A_k(\xi) \qquad \text{with} \qquad \|\Delta A_k(\xi)\| \leq \frac{1}{2}\:
|A_k(0)|\:. \]
Now~$\Delta A$ can be treated as a perturbation. Using that
the deviation of the eigenvalues is bounded by the sup-norm of the perturbation
 (see~\cite[Chapter~Two, \S 1]{kato}), we obtain the following estimate for the local trace,
\[ | A_k(\xi) | \geq \frac{1}{2}\: |A_k(0)|\:. \]
Hence integrating over the ball~\eqref{xireg}, we obtain the inequality
\[ \T_k \geq \frac{|A_k(0)|^2}{4}\: \frac{\pi^2}{2}\: \left(
8 n\, {\mbox{diam}}(\hat{K}) \right)^{-4} , \]
giving us the desired a-priori bound for the spectral weight~$|A_k(0)|$.
Again using that~$A_k(0) = P_k(0)^2$, it follows that
the spectral weight of~$P_k(0)$ is uniformly bounded, proving~\eqref{contcond}.
\QED

\Thanks{{\em{Acknowledgments:}} I would like to thank Andreas Grotz, Joel Smoller,
Daniela Schiefen\-eder and the referee for helpful comments on the manuscript.}


\def\dbar{\leavevmode\hbox to 0pt{\hskip.2ex \accent"16\hss}d}
\providecommand{\bysame}{\leavevmode\hbox to3em{\hrulefill}\thinspace}
\providecommand{\MR}{\relax\ifhmode\unskip\space\fi MR }
\providecommand{\MRhref}[2]{%
  \href{http://www.ams.org/mathscinet-getitem?mr=#1}{#2}
}
\providecommand{\href}[2]{#2}

\end{document}